\documentclass{aa}
\usepackage{txfonts}
\usepackage{epsfig}
\usepackage{rotating}

\usepackage{natbib}

\bibpunct{(}{)}{;}{a}{}{,}

\usepackage{times}
\usepackage{booktabs}
\usepackage{longtable}

\usepackage{color}

\usepackage[utf8]{inputenc}

\begin{document}

   \title{Spin period evolution of GX~1+4}

   \author{A. Gonz\'{a}lez-Gal\'{a}n\inst{1}
   	   \and
	   E. Kuulkers\inst{2}
	   \and
	   P. Kretschmar\inst{2}
	   \and
	   S. Larsson\inst{3,4}
	   \and
	   K. Postnov\inst{5}
	   \and 
	   A. Kochetkova\inst{5}	
	   \and
	   M.H. Finger\inst{6,7}}
	   
   \authorrunning{A. Gonz\'{a}lez-Gal\'{a}n}
   \titlerunning{Spin period evolution of GX~1+4}

   \offprints{A. Gonz\'{a}lez-Gal\'{a}n}

   \institute{Departamento de F\'isica, Ingenier\'ia de Sistemas y Teor\'ia de la Se\~nal, Universidad de Alicante, P.O.~Box 99 E-03080 Alicante, Spain
   \newline   
   \email{anagonzalez@ua.es}
	\and European Space Astronomy Centre (ESA/ESAC), Science Operations Department, Villanueva de la Ca\~nada (Madrid), Spain
	\and Department of Astronomy, Stockholm University, SE-106 91 Stockholm, Sweden
	\and The Oskar Klein Centre for Cosmoparticle Physics, AlbaNova, SE-106 91 Stockholm, Sweden
	\and Sternberg Astronomical Institute, Moscow MV Lomonosov State University, 119992 Moscow, Russia
	\and National Space Science and Technology Center, 320 Sparkman Drive, Huntsville, AL 35805, USA
	\and Universities Space Research Association, 6767 Old Madison Pike, Suite 450, Huntsville, AL 35806, USA}

   \date{Received; accepted}

  \abstract
   {}
   {We aim both to complement the existing data on the spin history of the peculiar accreting X-ray pulsar GX~1+4 with more past and current data from BeppoSAX, INTEGRAL, and Fermi and to interpret the evolution in the framework of accretion theory.}
   {We used source light curves obtained from BeppoSAX/WFC and INTEGRAL/ISGRI to derive pulse periods using an epoch-folding analysis. Fermi/GBM data were analysed by fitting a constant plus a Fourier expansion to background-subtracted rates, and maximizing the $Y_2$ statistic. We completed the sample with hard X-ray light curves from Swift/BAT. The data were checked for correlations between flux and changes of the pulsar spin on different timescales.}
   {The spin-down of the pulsar continues with a constant change in frequency, 
   i.e., an apparently accelerating change in the period. Over the past three 
   decades, the pulse period has increased by about $\sim$50\%.
   Short-term fluctuations on top of this long-term trend do show
    anti-correlation with the source flux.
    Possible explanations of the observed long-term frequency and its dependence on flux are discussed. }
   {}


   \maketitle
%

\parindent=0pt

\setlongtables

\section{Introduction}

Accreting X-ray pulsars are highly magnetized neutron stars in a binary system, which accrete matter from their companion star. The mass transfer can take place via Roche-Lobe overflow for low mass X-ray binaries (LMXBs), strong stellar winds for giant stars in high mass X-ray binaries (HMXBs), or the Be emission mechanism for Be X-ray binaries. These accreting pulsars radiate predominantly in the X-ray band, and the radiation is modulated by the stellar rotation of the pulsar. For a review of this subject, see e.g., \citet{nagase89}.

Since the discovery of the first accreting X-ray pulsars, Cen~X-3 \citep{Giacconi}, GX~1+4 \citep{GX71Lewin}, and Her~X-1 \citep{Tanan}, the pulse periods of well-established accreting X-ray pulsars have been monitored more or less regularly with a wide variety of high-energy observatories. Early observations mainly found sources being spun up, which was easily explained by the acceleration through the angular momentum of accreted matter. Further investigations have demonstrated a wide variety in the pulse-period evolution, with some sources best described by a random walk, others showing clear secular changes. Various models \citep[e.g.,][]{GhoshLambModel, wang87, LovelaceModel} have been proposed to explain the observed behaviour. 

GX~1+4 was discovered in 1970 by a balloon X-ray observation at energies above 15 keV showing pulsations with a period of about two minutes \citep{GX71Lewin}. During the 1970s, it was one of the brightest X-ray sources in the Galactic centre region and was globally spinning up strongly \citep[e.g.,][]{GX81Doty, GX81Warwick, GX83White, GX89Mc}. In the early 1980s the source went through a low state in X-ray flux and remained undetectable, because at least two orders of magnitude below the previously observed levels \citep{GX83IAU, GX88Mukai}. When the source was detected again it had undergone a torque reversal \citep{GX88Mak}, and ever since it has been generally spinning down strongly, see Appendix~\ref{spinluminosity}.

The optical counterpart of GX~1+4 was discovered by \cite{GX73Glass} as the infrared source V2116~Oph. The optical composite emission spectrum of the proposed counterpart indicated that the object was almost certainly a binary system, consisting of a symbiotic red giant and a much hotter source \citep{GX77Davidsen}. \citet{GX97Roche} confirm the optical  companion to be V2116~Oph. More recently, infrared observations have indicated a mass of about 1.2 $M_\odot$ for the M giant star (assuming a mass of about 1.35 M$_\odot$ for the neutron star), implying that the M giant star is a first ascent giant that does not fill its Roche Lobe \citep{GX06Hinkle}. Therefore, GX~1+4 is a LMXB that is capturing the stellar wind of its M6III companion \citep{GX97Roche,GX06Hinkle}. GX~1+4 is the first and the prototype of the small but growing subclass of accreting X-ray pulsars called symbiotic X-ray binaries (SyXB), by analogy with symbiotic stars, in which a white dwarf accretes from the wind of an M-type giant companion \citep[e.g.,][]{Masetti2006_SyXB, Masetti2007_IGRJ_SyXB, GX08Corbet}.

\cite{GX86Cut} proposed an orbital period of about 304 days based on variations in the pulse period of the neutron star of GX~1+4 during the spin-up phase in the 1970s. Other authors \citep[e.g.,][]{GXPereira99,GX2000Braga} have supported this using observations from 1991 to 1998, when the source was already in its long-term spin-down phase (see also Fig.~\ref{GXhistdata}). More recently, infrared observations have shown a 1161-day period single-line spectroscopic binary orbit, which excludes the 304-day period as the orbital period \citep{GX06Hinkle}. GX~1+4's X-ray light curves show strong variability on a timescale of days to years but with no modulation on either the optical 1161-day orbital period or the previously reported 304-day X-ray \citep[e.g.,][]{GX05Naik, GX08Corbet}.

Based on X-ray and infrared observations, \citet{GX97Roche} have constrained the distance range to the system as between 3 and 15 kpc, depending on the evolutionary state of the red giant. 
For a first-ascent giant branch star \citep{GX97Roche, GX06Hinkle}, the estimates have narrowed down to 3-6 kpc \citep{GX97Roche}.
The distance estimate by \citet{GX06Hinkle}, 4.3 kpc, is consistent with this. 
Thus, in spite of its position in the sky (see, e.g., \citet{GX95Pred}), GX~1+4/V2116~Oph is clearly not an object associated with the centre of the Milky Way \citep{GX06Hinkle}.
\citet{GX06Hinkle}, however, do not provide any uncertainty on the distance, and an uncertainty of 0.1 kpc, based on the precision of their estimate, seems to be too precise. They do provide an uncertainty in the effective temperature of the giant star ($\pm$200~K) and its radius (+42,$-$30~R$_{\odot}$), which would give a range in the absolute luminosity, and therefore, an uncertainty in the distance may be estimated, while taking the interstellar extinction into account. Estimates of the latter, in terms of $E(B-V)$, have been reported by various authors: 1.62$\pm$0.19 \citep{GX97Roche}, 1.7$\pm$0.4 \citep{GX77Davidsen}, 2.1$\pm$0.1 \citep{GX97Jablon}, 2.30$\pm$0.06 \citep{GX96Shahbaz}.\footnote{The observed optical extinction is mainly attributed to interstellar dust, while the variable X-ray absorption is possibly due to the wind of the giant star, see \citet{GX97Roche}.}
This rather wide range in extinction has a strong effect on the uncertainty in the distance, i.e., about a few kpc. 
With the mean observed V-band magnitude, V$=18.40 \pm 0.03$~mag, $E(B-V)=1.62 \pm 0.19$ or, equivalently, $A_{\rm V}=5.0 \pm 0.6$ \citet{GX97Roche}, and assuming an absolute magnitude $M_{\rm V}$ of about $-$0.25 for an M6III star \citep{GX90The}, we derive a distance of 5.4 $\pm$ 1.6~kpc.
A higher reddening value, for example, of $E(B-V)=2.30 \pm 0.06$ or, equivalently, $A_{\rm V}=7.1 \pm 0.2$ \citep{GX96Shahbaz} would lead to a distance of 2 $\pm$ 0.2~kpc.
We, therefore, conclude that \citet{GX97Roche} give a more realistic estimate than \citet{GX06Hinkle} and that the uncertainty on the distance is about 1.5~kpc. In this paper we adhere to the distance estimate of 4.3~kpc. 

Currently, there is no well established value for the magnetic field of \mbox{GX~1+4}. Assuming the standard accretion-disc theory \citep{ghoshtransition,ghoshtorques,wang87}, the magnetic field has been estimated to be $ B\sim  10^{13}-10^{14}$~G \citep[e.g.,][]{GX89Dotani,GX91Mony,GX04Cui}, which is among the largest measured for any accreting X-ray pulsar.  There have been marginal reports of cyclotron scattering resonance reatures (CRSFs) in the X-ray spectra, which points to a value of $ B \sim 10^{12} $~G for the magnetic field, i.e., up to two orders lower \citep[e.g.,][]{GX05Rea, GX07Ferrigno}.

GX~1+4's unusual long-term spin behaviour has attracted considerable interest for many years. Studying the pulse-period evolution in an accreting X-ray pulsar and relating it to, e.g., the luminosity changes, allows testing theoretical models and gaining insight into the interaction between the pulsar's magnetosphere and the accreted matter. In this paper we extend the investigation of the spin-period history of GX~1+4 with new observations obtained by BeppoSAX, INTEGRAL, and Fermi. The available measurements of the pulse period of GX~1+4 span a period of about 40 years. This gives us a unique insight in the pulse period evolution of this SyXB. Apart from GX~1+4 still spinning down overall, we find irregular trends on top of this evolution. We correlate these features with the high-energy flux of the system using the same observations as well as those obtained with Swift, and provide an explanation for the spin history seen.



\section{Observations}
\label{Observations}

\subsection{CGRO}

The Compton Gamma-Ray Observatory, CGRO, was a NASA mission launched in April 1991 \citep{CGRO93} and operative until June 2000 \citep{CGRO06} . The Burst And Transient Source Experiment, BATSE, onboard CGRO was a sensitive all-sky instrument that consisted of eight uncollimated Na I scintillation detectors at the corners of the spacecraft. It covered a broad energy range from 15~keV to 100~MeV. Each detector module contained a large-area detector (LAD) optimized for sensitivity and directional response and a spectroscopy detector (SD) optimized for broad energy coverage and energy resolution \citep{BATSE92}.

BATSE monitored pulse frequencies and X-ray pulsed fluxes for roughly half of the known X-ray pulsars \citep{BATSEmonitoringNelson}. GX~1+4 results from this survey are publicy available in the form of pulse source histories covering the energy range from 20~keV to 50~keV and daily frequencies from April 1994 to May 1997.\footnote{{See \tt \tiny ftp://legacy.gsfc.nasa.gov/compton/data/batse/\ pulsar/histories/}.}

\subsection{BeppoSAX}

The X-ray satellite ``Satellite per Astronomia X'', BeppoSAX, was an Italian/Dutch mission launched in April 1996 and operated until April 2002, then deorbited in April 2003. It covered more than three decades in energy (about 0.1 to 300 keV) with relatively good energy resolution, and provided imaging capabilities in the range of 0.1-10 keV. Together with the Wide-Field Camera's (WFCs), the broad-band Narrow-Field Instruments (NFIs) provided the opportunity to study the broad-band behaviour of several classes of X-ray sources \citep{BeppoSAX3}.

The WFCs \citep{BeppoSAX4} were two identical coded-aperture instruments onboard BeppoSAX. The field of view was 40$\degr$$\times$40$\degr$ full width zero response (FWZR), the angular resolution 5$\arcmin$ full width half maximum (FWHM) and the source-location accuracy was generally better than 1$\arcmin$ (99\%\ confidence).  The detectors were sensitive to the energy range 2 to 28\,keV.
The WFCs pointed in opposite directions with respect to each other and perpendicular to the NFIs. The pointing directions of the WFCs were usually governed by the observations of the NFIs. However, a few dedicated campaigns on the Galactic bulge region were performed by the WFCs \citep[see, e.g.,][]{BeppoSAX5}. We analysed WFC data obtained from August 1998 to August 2000.

\subsection{INTEGRAL}

INTEGRAL \citep{Integral} is an ESA scientific mission that was launched in October 2002. It is dedicated to spectroscopy and imaging of celestial \begin{math}\gamma\end{math}-ray sources in the energy range between 15 keV and 10 MeV with simultaneous monitoring in the X-ray and optical energy ranges. Several instruments are on board: SPectrometer onboard INTEGRAL (SPI), Imager on Board INTEGRAL Satellite (IBIS), Joint European Monitor X-rays (JEM-X) and Optical Monitoring Camera (OMC). The former three instruments all collect photons through wide-field coded masks.

IBIS \citep{IBIS} comprises two detector planes. The field of view is 29$\degr$$\times$29$\degr$ (FWZR) and the angular resolution 12$\arcmin$ (FWHM). In this paper we only used data collected with one of its detectors: the INTEGRAL Soft Gamma-Ray Imager (ISGRI). It is sensitive in the $\simeq$15\,keV to 1\,MeV range \citep{ISGRI}.

The INTEGRAL Galactic bulge monitoring programme started in February 2005 and was initiated to monitor the Galactic bulge region on a regular basis in mainly the hard X-ray band. One complete hexagonal dither pattern (7 pointings of $\simeq1800$ s each) is performed during each INTEGRAL orbit around the Earth, (roughly every 3 days), whenever the bulge region is visible by INTEGRAL: twice per year for a total period of about four months \citep[see][]{galbulge}.

\begin{table}
\caption{\label{OP} INTEGRAL observation periods (OP)}
\begin{center}
\begin{tabular}{lllll}
OP & Start Date & End Date & MJD & Effective\\
 & \multicolumn{2}{c}{(UT)} & & Exposure\\
\hline
1 & 2005/02/18 & 2005/04/19 & 53419-53479 & 258 ks  \\
2 & 2005/08/16 & 2005/10/26 & 53598-53669 & 377 ks \\
3 & 2006/08/9  & 2006/04/21 & 53775-53846 & 276 ks \\
4 & 2006/08/13 & 2006/10/23 & 53963-54031 & 461 ks \\
5 & 2007/02/15 & 2007/04/21 & 54146-54211 &  853 ks\\
6 & 2007/08/19 & 2007/10/14 & 54331-54387 & 599 ks\\
7 & 2008/02/11 & 2008/04/20 & 54507-54576 &  225 ks\\
\hline
\end{tabular}
\end{center}
\end{table}

We analysed the ISGRI data from observations of the Galactic bulge monitoring programme between February 2005 and April 2008. These observations span several periods, see Table~\ref{OP}. Each time span is referred to as observation period, or OP for short.

\subsection{Swift}

Swift is a NASA mission that was launched in November 2004 \citep{swift1}. The Burst Alert Telescope \citep[BAT;][]{swift2} onboard Swift is a coded-aperture imager with a very wide field of view of about 2 steradians, which operates in the 15--150\,keV band. The BAT angular resolution is 22$\arcmin$ (FWHM).

The BAT continually monitors the sky with more than about 70\%\ of the sky observed on a daily basis. Results from this survey are publicly available in the form of light curves covering the 15--50\,keV energy band on two timescales: a single Swift pointing ($\simeq$20\,min) and the weighted average for each day.\footnote{{\tt \tiny http://swift.gsfc.nasa.gov/docs/swift/results/\ transients/index.html}.} We used the daily average light curve for GX~1+4 to study its long-term hard X-ray flux behaviour. 

\subsection{Fermi}

The Fermi Gamma-ray Space Telescope, formerly Gamma-ray Large Area Space Telescope (GLAST), is a NASA mission that was launched in June 2008. It has two instruments on board: the Large Area Telescope (LAT) and the Gamma-ray Burst Monitor (GBM). Fermi is dedicated to measuring the cosmic gamma-ray flux in the energy range 20 MeV to $>$ 300 GeV, with supporting measurements for gamma-ray bursts \citep{Fermi}. 

Since 2008 August 12 GX~1+4 has been continuously monitored by the GBM \citep{FermiMeegan}. The GBM is an all-sky instrument sensitive to X-rays and gamma rays with energies between $\sim$8~keV and $\sim$40~MeV. Timing analysis is carried out with channels 1 and 2 of the NaI detector {\sc CTIME} data (12--50~keV, 0.256~s time resolution).

We analysed GBM data from August 2008 to February 2010.

\section{Data analysis}

For INTEGRAL/ISGRI, the data were reduced using version 7 of the INTEGRAL off-line analysis software ({\sc OSA 7}), distributed by the INTEGRAL Science Data Centre \citep[ISDC;][]{ISDC}. Each data set in a revolution (about 13 ks long) has been analysed in the energy range between 20 keV and 40 keV, following the steps described in the IBIS Analysis User Manual (version 6.0). We used the \emph{ii\_light} tool to obtain light curves with a time bin of 10~s and barycentric correction was applied. The source BeppoSAX/WFC flux was reconstructed in the 2--25\,keV bandpass with a time resolution of 2\,s.

Period determinations for both the INTEGRAL/ISGRI and BeppoSAX/WFC data have been done for segments of the light curve using an epoch-folding analysis \citep{1996Larsson}. The data for such a segment is folded at a number of different test periods, and for each period, the $\chi^2$ over the resulting pulse profile is computed. A best-fit period is determined by fitting a template function describing how $\chi^2$ should vary with a test period. The template function takes the time sampling of the data into account and, in an iterative procedure, the pulse shape of the oscillation. Uncertainties are estimated by Monte Carlo simulations. A set of synthetic pulse light curves with the same time sampling and noise level as the data are created and analysed. The distribution of determined periods for these simulations is used as a measure of period uncertainty. Pulse period determination for INTEGRAL/ISGRI data was possible for most hexagonal pattern observations. From a total of 128 hexagonal dither patterns it was possible to obtain 121 light curves (in 7 of these 128 observations the source was below the detection limit of the instrument) and to derive 93 pulse periods, which is $\sim$77\% of the detections. For 28 detections it was not possible to derive the pulse period because the source detection significance was lower than 14 and no clear maximum in the $\chi^2$ distribution was found owing to the noise.

The analysis of the Fermi/GBM data is complicated by Fermi's continuously changing orientation. All intervals of {\sc CTIME} data from the 12 NaI detectors are selected for analysis where the high voltage is on, excluding those containing high-voltage transients, phosphorescence events, rapid spacecraft slews, South Atlantic Anomaly induced transients, electron precipitation events, and gamma-ray bursts. Source pulses are then separated from the background by fitting the rates in all detectors with a background model and subtracting the best-fit model. This model includes bright sources and their changing detector responses (including Earth occultation steps), along with quadratic spline functions that account for the remaining long-term background trends. The spline models have statistical constraints on the changes in second derivative between spline segments to control the model stiffness. These fits are made jointly across detectors (with  common bright source fluxes) but separately for each channel of the {\sc CTIME} data. The residuals are then summed over detectors with time-dependent weights that are proportional to the predicted (phase-averaged) count rates from the source. Short intervals (900s) of these combined residuals are then fitted with a constant plus a Fourier expansion to determine a pulse profile. The profiles are divided into four-day intervals, and the pulse  frequency and mean profile are then determined in each interval with a search of pulse frequency for the maximum of the $Y_n$ ($n=2$) statistic \citep{FermiMark}. $Y_n$ was formulated to find a pulse frequency from a series of pulse profiles, each represented by a finite Fourier expansion, so it accounts for possible frequency-dependent non-Poison noise.

We did not apply Doppler corrections to the data, since the projected semi-major axis has not been measured with X-ray observations. The latter is mainly due to the high level of torque variability at low frequencies.

\section{Pulse period evolution}
\label{Results}

\begin{figure}
\begin{center}
\epsfig{file=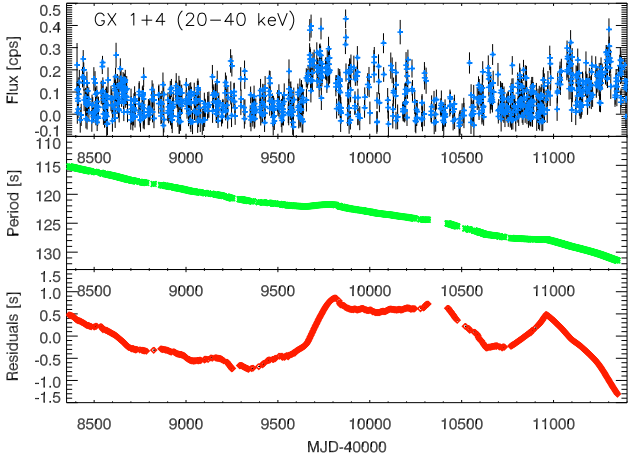, width=6cm,angle=90}
\end{center}
\caption{\label{GX_BATSE_old} \textit{Top:} GX~1+4 CGRO/BATSE daily light curve in the energy range 20-40 keV. \textit{Middle:} Pulse periods derived from CGRO/BATSE data, with the period increasing from top to bottom. \textit{Bottom:} Residuals of the periods from a linear fit.}
\end{figure}

\begin{figure}
\begin{center}
\epsfig{file=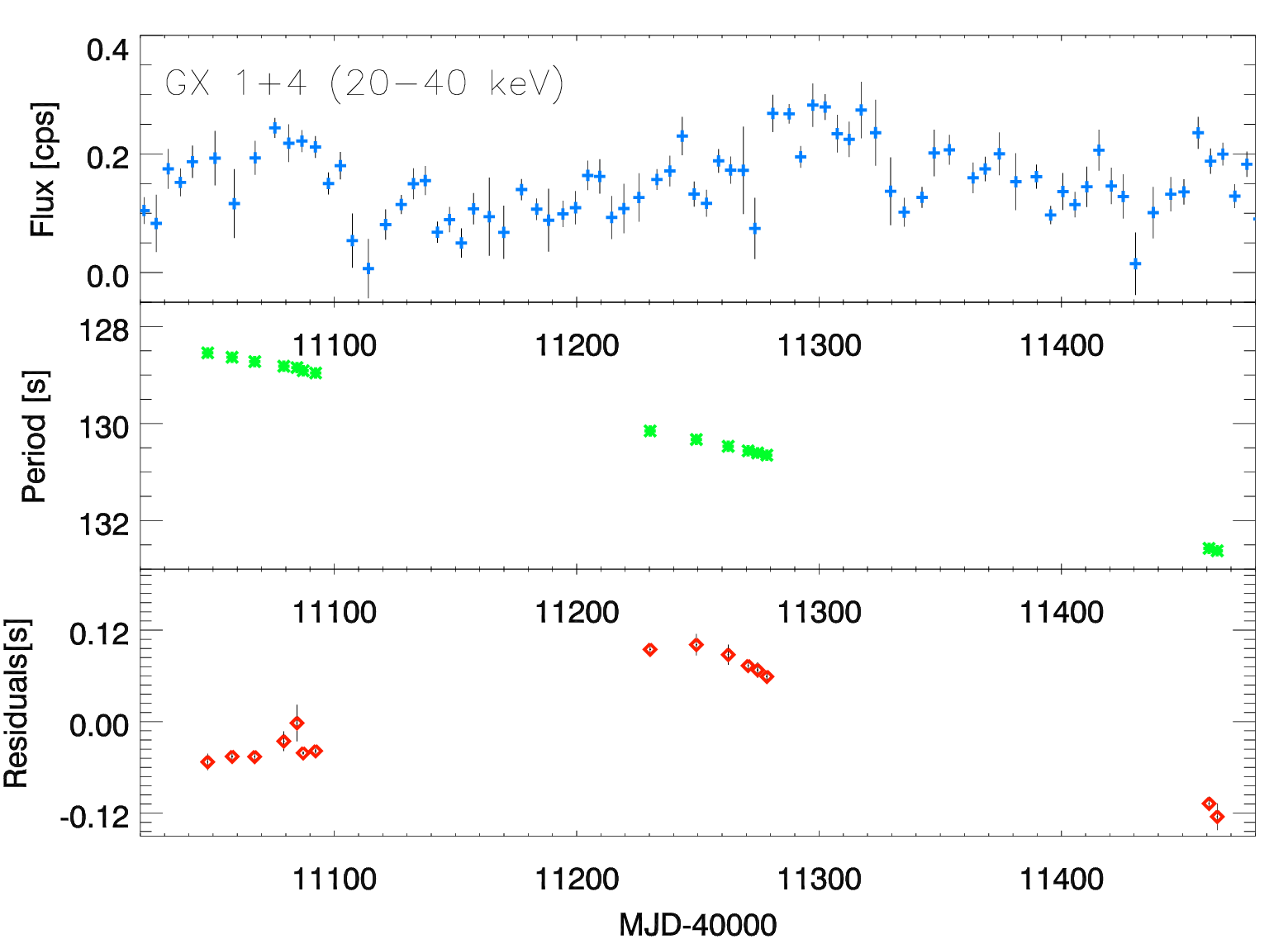, height=6cm}
\end{center}
\caption{\label{GX_BATSE_SAX} \textit{Top:} GX~1+4 CGRO/BATSE daily light curve in the energy range 20-40 keV. \textit{Middle:} Pulse periods derived from BeppoSAX/WFC data. Note that the period increases from top to bottom. \textit{Bottom:} Residuals of the periods from a linear fit.}
\end{figure}

\begin{figure}
\begin{center}
\epsfig{file=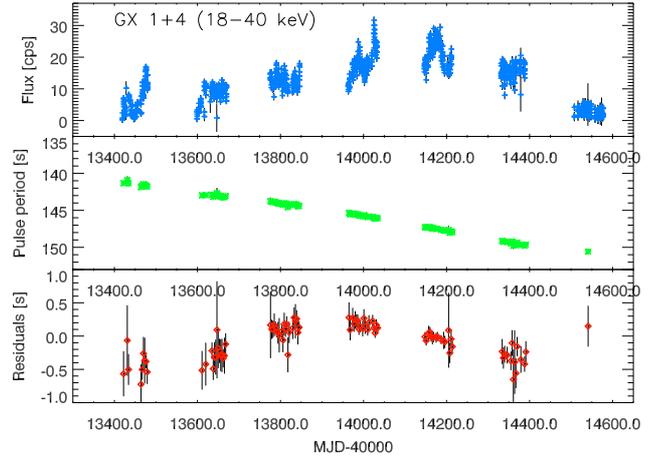, height=6cm}
\end{center}
\caption{\label{IntegralResults} \textit{Top:} GX~1+4 INTEGRAL/ISGRI average flux per pointing in the energy range 18 to 40 keV. \textit{Middle:} Pulse periods derived from INTEGRAL/ISGRI data.  Note that the period increases from top to bottom. \textit{Bottom:} Residuals from the pulse periods using a linear fit.}
\end{figure}

\begin{figure}
\begin{center}
\epsfig{file=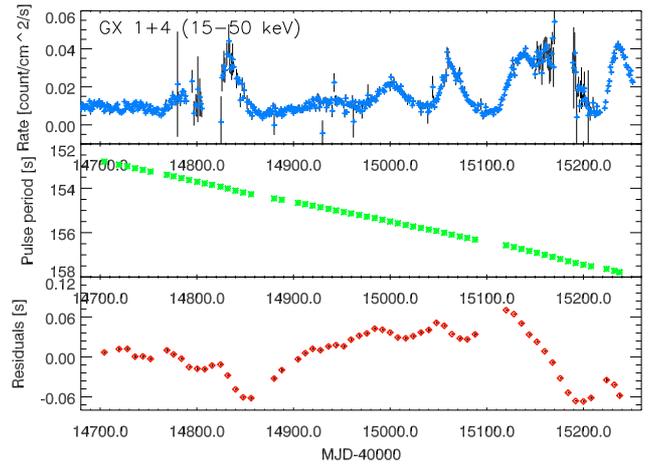, height=6cm}
\end{center}
\caption{\label{GXFermi} \textit{Top:} GX~1+4 Swift/BAT daily averaged light curve in the energy range 15--50 keV. \textit{Middle:} Pulse periods derived from Fermi/GBM data, with the period increasing from top to bottom. \textit{Bottom:} Residuals from the pulse periods using a linear fit.}
\end{figure}

\begin{figure}
\begin{center}
\epsfig{file=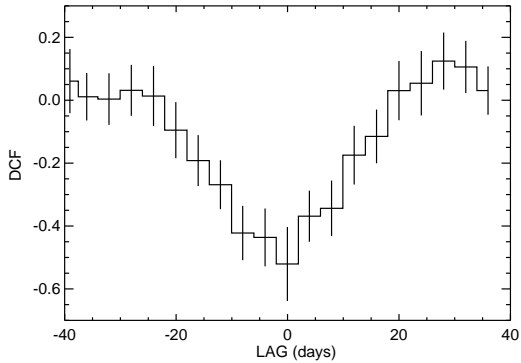, height=5cm}
\end{center}
\caption{\label{CC}%
The discrete correlation function (DCF), i.e., the correlation coefficient between the
15--50~keV X-ray flux and the pulse period change $\dot\nu$, as function of the
time lag. The minimum near zero lag implies an anti-correlation between the pulse frequency derivative and X-ray flux.}
\end{figure}

\subsection{CGRO}
The CGRO/BATSE data can be found in Fig.~\ref{GX_BATSE_old}.  The period increases in time with an almost linear trend, but with some fluctuations. A linear fit to these data resulted in a slope of (5.61 $\pm$ 0.03) $\times$ $10^{-8}$ s/s with a Pearson $r^2$ value of 0.994. The deviations from this linear fit are shown in the bottom panel of Fig.~\ref{GX_BATSE_old}. 
In this figure it is possible to observe a deviation from the linear fit of the spin period evolution around MJD 49700, where the neutron star spin-down rate decreases in coincidence with an increase in the X-ray flux. Overall, the deviations in the linear fit shown in the bottom panel of this figure are visually comparable to the variations in the X-ray flux shown in the upper panel of the same figure.

\subsection{BeppoSAX}
The BeppoSAX/WFC pulse periods are shown in Fig.~\ref{GX_BATSE_SAX}. The pulse period continues to increase with time. A linear fit to the pulse period as a function of time results in a slope of (1.113 $\pm$ 0.017) $\times$ $10^{-7}$~s/s with a Pearson $r^2$ value of 0.998. The residuals from this linear fit are shown in the bottom panel of Fig.~\ref{GX_BATSE_SAX}. The CGRO/BATSE X-ray pulsed flux contemporaneous to the WFC data is shown in the top panel of this figure.
The sparsity of the pulse period measurements means we cannot see any clear correlations between the X-ray flux and the spin-period evolution in this data set.

\subsection{INTEGRAL}
The results using the INTEGRAL/ISGRI data are shown in Fig.~\ref{IntegralResults}. The pulse period increases with time with an almost linear trend. We find a slope of (1.031 $\pm$ 0.007) $\times$ $10^{-7}$ s/s with a Pearson $r^2$ value of 0.995. As can be seen in the bottom panel of Fig.~\ref{IntegralResults}, the last point at MJD 54541 is only a marginal detection of the pulse period. Any linear fit not taking this last data point into account does not lead to a significantly different slope. The deviations from the linear fit not taking it into account the last data point are shown in the bottom of Fig.~\ref{IntegralResults}. In this data set there is a maximum for the deviations of the spin-period linear fit around MJD 53900, followed by a maximum for the X-ray flux around MJD 54100.

\begin{figure*}
\begin{center}
\epsfig{file=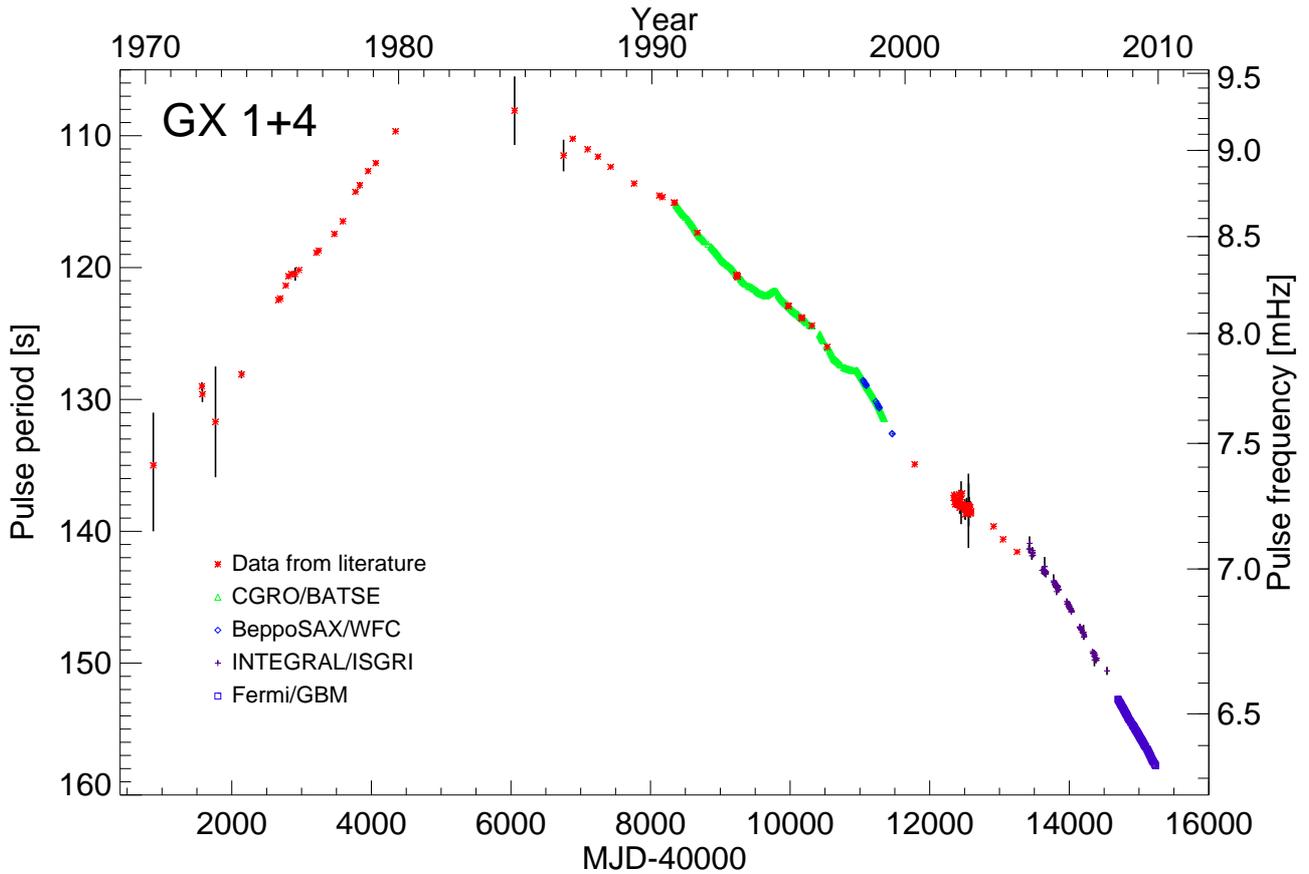, height=12cm}
\end{center}
\caption{\label{GXhistdata}The long-term pulse period evolution of GX~1+4, incorporating data from the literature and results of this work. See Table~\ref{GXPulseHistory} for data and references.}
\end{figure*}

\subsection{Fermi}
\label{Results:Fermi}
The pulse period results using the Fermi/GBM data can be seen in the middle panel of Fig.~\ref{GXFermi}. Again, the period evolution for GX~1+4 follows an almost linear trend with a slope of (1.0697$\pm$0.0002) $\times$ $10^{-7}$ s/s and a Pearson coefficient of $r^{2}=0.999$. The evolution of the Swift/BAT hard X-ray flux is shown at the top of this figure.

\subsection{X-ray flux versus spin period}

Theoretical models predict certain correlations between X-ray fluxes and pulse period evolution.
For example, the complete and continuous data series in Fig.~\ref{GX_BATSE_old} led
\cite{GX97Chack} and \cite{Nelson} to demonstrate for the first time a negative correlation between spin-up rate $\dot\nu$ and X-ray luminosity. Motivated by this, we searched for correlations in the data shown in Fig.~\ref{GXFermi} which represents the most complete and continuous data series of period and fluxes.
We calculated the discrete correlation function, DCF \citep[see][]{CCek,CCp}, between $\dot \nu $, estimated over eight day bins, and the one-day binned Swift/BAT 15-50 keV flux. The DCF for the full MJD 54700--55240 time range (Fig.~\ref{CC}) shows a strong negative correlation at zero lag ($-1.3 \pm 2.6$~days). Since the main uncertainty of the DCF is random correlation with erratic X-ray flickering in the light curve, we divided the data into four sections and computed the DCF for each. Strong anticorrelation ($\sim -0.6$) was seen in the last two of these, and it was weaker ($\sim -0.3$ to $-0.4$) in the first two sections. In all four the DCF minimum for lags between $-40$ and $+40$~days occurred close to, and was consistent with, no lag. 

Comparing the instantaneous spin frequency derivative derived from the
Fermi/GBM data with the flux simultaneously measured by the Swift/BAT monitor
we find a dependence $-\dot\nu\propto F_x^{(0.30\pm0.07)}$. This is
in line with the correlation of instantaneous spin-down torque with X-ray flux discovered by BATSE \citep{GX97Chack}. 

\subsection{Long-term pulse period evolution}
\label{longtermevolution}

The long-term pulse period evolution of  \mbox{GX~1+4} is shown in Fig.~\ref{GXhistdata}. We combined all measurements that we are aware of from the literature with the pulse periods determined from BeppoSAX/WFC, INTEGRAL/ISGRI, and Fermi/GBM data as presented in the previous subsection. The main features in Fig.~\ref{GXhistdata} are the well known switch from a strong spin-up to a spin-down trend in the 1980s and a continued spin-down slowly increasing in average $\dot{P}$ over the years. The spin-down rate observed  around 2004 was $\dot{P}\sim 10^{-8}$~s/s, while the spin-down trend observed in this work with INTEGRAL/ISGRI (2005-2008) and with Fermi/GBM (2009-2010) is $\dot{P}\sim$ $10^{-7}$~s/s, taking the spin-down measure in frequencies, the global evolution since the start of spin-down is described very well by a linear trend of \mbox{$-0.1177(3)$~mHZ/y} as it is shown in the upper panel of Fig.~\ref{PetersFigure}. 

On top of these spin-down trends, irregularities are seen that have sometimes been proposed to correlate with the binary orbit \citep[e.g.,][]{GXPereira99, GX2000Braga}. (For a more detailed description see Section~\ref{accretionmodels} and Appendix~\ref{spinluminosity}). These irregularities are reflected in the middle panel of Fig.~\ref{PetersFigure}, but we note also that other apparent trend changes occur at times far from the predicted perigee passages.

\section{Discussion}
\subsection{A retrograde disc in GX~1+4?}
\label{accretionmodels}

When GX~1+4 was discovered, the pulsar was spinning up, and the spin-up rate $\dot\nu$ was apparently correlated with the observed X-ray flux, which is usually interpreted as a direct measure of X-ray luminosity, hence mass accretion rate $\dot{M}$ \citep[e.g.,][]{GX81Doty, GX82Ric}. This supported the idea that the period decrease is produced by accretion torques of a prograde disc around the neutron star \citep[e.g.,][]{ghoshtorques, wang87}.

Around 1983--1984 GX~1+4 entered a low X-ray luminosity state \citep{GX83IAU,GX88Mukai}, suggesting a large reduction in the mass accretion rate, and it started to spin down.
\cite{GX88Mak} were the first to propose for GX~1+4 that a retrograde disc is formed around the neutron star by matter captured from the stellar wind of the M-giant companion of the pulsar, to explain the observed spin-down. Since then, this retrograde disc hypothesis has been supported by several authors \citep[e.g.,][]{GX89Dotani, GX97Chack, Nelson} for different reasons.
One of the arguments in favour of the retrograde disc hypothesis is related to the magnetic field. To explain the spin-down of a pulsar accreting from a prograde standard disc \citep[see e.g.,][]{GhoshLambModel, ghoshtransition, ghoshtorques, wang87}, the neutron star must be rotating very near its equilibrium spin period. The pulsar should then be in a quasi-equilibrium state, and GX~1+4 consequently has the strongest known magnetic field of any neutron star ($B\sim10^{13}-10^{14}$~G \citep[e.g.,][]{GX89Dotani, GX91Mony, GX93Green, GX04Cui}. The retrograde disc scenario  would eliminate the need for such an unusually strong magnetic field, but the pulsar would necesarily be far from its equilibrium spin period \citep[e.g.,][]{GX88Mak, GX89Dotani, GX97Chack, Nelson}.

\begin{figure*}
\sidecaption
\includegraphics[width=12cm]{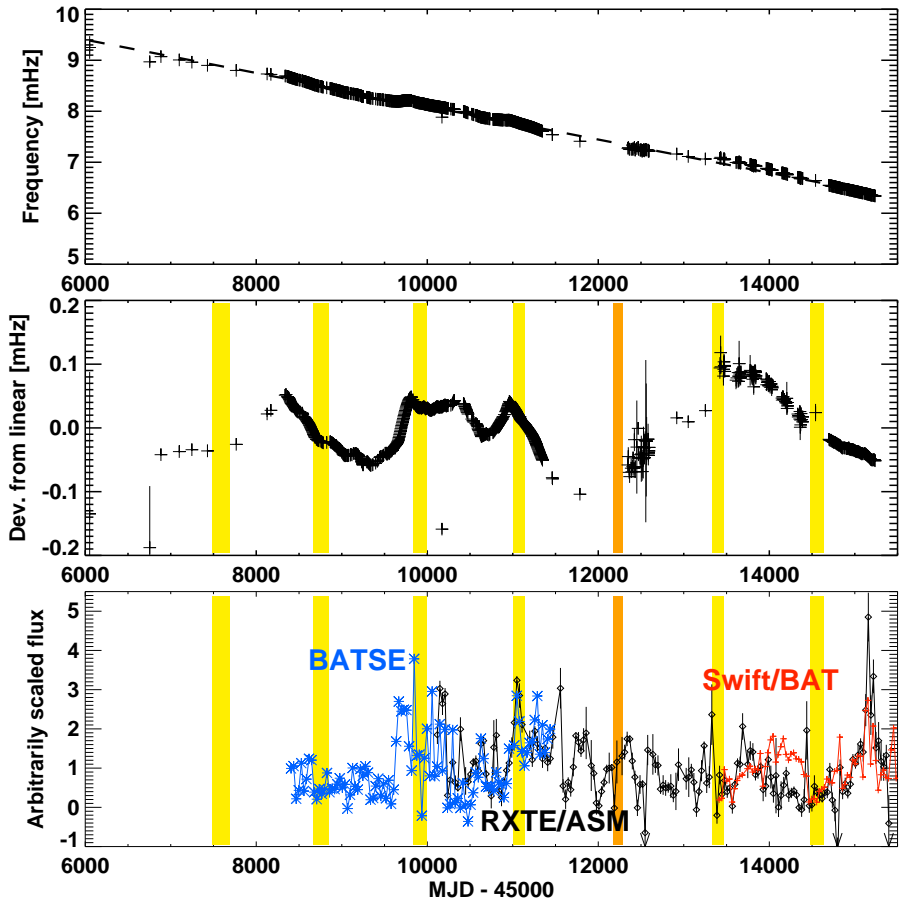}
\caption{\textit{Top:} Long-term spin-down of GX~1+4 expressed in frequencies. \textit{Middle:} Residuals from the linear fit to the frequencies. The orange shaded area indicates the perigee passage time, including uncertainties, given by \protect{\citet{GX06Hinkle}}. Yellow shaded areas indicate the perigee passage intervals extrapolated
from this ephemeris. \textit{Bottom:} Observed X-ray fluxes in arbitrary scaling to the same mean by BATSE, RXTE/ASM, and Swift/BAT, averaged over 30-day intervals.
}
\label{PetersFigure}
\end{figure*}

The negative correlation of spin-up rate $\dot\nu$ and X-ray flux found in BATSE observations in the 1990s \citep{GX97Chack,GX97Paul} and in this work is also not consistent with the standard disc accretion model by \cite{ghoshtorques}, which predicts higher spin-up rates for higher X-ray luminosities. However, it is consistent with a retrograde disc around the neutron star, taking angular momentum off the pulsar, so that higher spin-down rates for higher X-ray luminosities are expected.
However, it remains a chief question how a retrograde disc could form and remain stable over such a long time period (about 30 years).

On the other hand, positive correlations between spin-up rate $\dot\nu$ and X-ray luminosity have also been found during the steady spin-down trend \citep[e.g.,][]{GX97Chack}. In addition to these positive correlations, during this spin-down steady trend some spin-up episodes related to bright flares, where the source has reached almost the same luminosity as it had in the 1970s, have taken place \citep[e.g.,][]{GX97Chack, GX07Ferrigno}. Again, it is not possible to explain this with accretion from a retrograde disc.

The presence of fast flickering has been interpreted as evidence of accretion discs around neutron stars in general \citep[e.g.,][]{horne94} and in particular, for GX~1+4 on the basis of such flickering in the optical light curves \citep[e.g.,][]{GX93Braga,GX97Jablon}. Meanwhile, the absence of this flickering when the system is faint has been interpreted as episodic interruptions of the accretion \citep{GX97Jablon}. Furthermore, coherent optical pulsations of two minutes found by \cite{GX97Jablon} in GX~1+4 have also been interpreted as a sign of X-ray reprocessing in an accretion disc \citep{Chester79} by several authors \citep[e.g.,][]{GX97Roche}. Moreover, \cite{GX97Roche} have carried out an optical emission-line diagnostic study of the optical and infrared spectra, which also suggested there is an accretion disc around the neutron star.

According to numerical simulations of mass accretion onto the neutron star in wind-fed systems, the accreted specific angular momentum could change sign in an erratic manner, which may lead to alternating spin-up and spin-down episodes \citep[e.g.,][]{Taam1988, Matsuda1991,2005fogli}. This phenomena, known as \textquotedblleft~flip-flop~\textquotedblright, has been used to explain the random walk of the spin period of sources like Vela~X-1 \citep[e.g.,][]{1984boyn}. However, it is important to note that these simulations apply to supergiant stars with powerful winds \citep[see e.g.][]{kudri2000}, and not to M-giant stars.\footnote{V2116~Oph is an M6-giant \citep{GX97Roche,GX06Hinkle}. Such stars have slow and strongly variable stellar winds \citep[see e.g.,][]{Crowley06}.}

Last but not least, the X-ray luminosity increase accompanied by a rather constant spin-down rate observed by, e.g., Ginga in 1987 \citep{GX90Sakao} implies (temporary) luminosity-independent spin-down behaviour. This cannot be understood in terms of either the prograde disc, or the retrograde one.

To conclude, ever since GX~1+4 entered the low-luminosity state around 1984, overall it kept spinning down with the general spin-down accompanied by spin-up episodes during flares. The presence of both positive and negative correlations between X-ray flux and spin-down rate, and episodes of no correlations between X-ray flux and spin-down rate, are not possible to explain either with standard disc accretion or with a retrograde disc. One cannot explain the long-term steady spin-down with an alternating disc either, because the neutron star tends to spin-down (except for short episodes). It can also not be explained by a retrograde disc, as the spin-down rate is increasing with time while the average X-ray luminosity\footnote{The luminosity value has been corrected to the distance of 4.3 kpc \citep{GX06Hinkle}, as are the luminosity values used hereafter.} is stable around $L_X\sim 10^{35}-10^{36}$~erg~s$^{-1}$.


\subsection{Quasi-spherical accretion in GX 1+4}
\label{quasisphe}

In this section we discuss how the long-term spin-down behaviour of GX 1+4 presented in Fig.~\ref{GXhistdata} can be explained by quasi-spherical accretion onto the neutron star from the stellar wind from the secondary companion.

In  wind-fed pulsars with long orbital periods,  accretion onto the neutron star can proceed quasi-spherically; i.e. an accretion disc around the neutron star magnetosphere cannot be formed at all. The formation of the disc depends on whether the specific angular momentum of matter $j_m$ near the magnetospheric radius $R_m$ is larger or smaller than the Keplerian value $j_K(R_m)=\sqrt{GMR_m}$. Assuming the specific angular momentum conservation, $j_m$ can be related to that of gravitationally captured stellar wind matter in the zone of bow shock at the Bondi radius $R_G=2GM/(V_w^2+V_{orb}^2)\sim 10^{12} v_7^{-2}$~cm (here $v_7$ is the relative stellar wind velocity in units of 100 km/s) , $j_m=j_w$. To within a numerical factor of order one, $j_w=(2\pi/P_{orb})R_G^2\approx 3 \times 10^{17} (P_{orb}/1000^d)^{-1}v_7^{-4}$, which can be smaller than $j_K(R_m)$ for typical magnetospheric radii\footnote{The value of the magnetospheric radius is determined by the pressure balance at the magnetospheric boundary, and generally can be notably different for disc and spherical accretion} $R_m=10^9$~cm, so the formation of quasi-spherical accretion flow is very likely. 

There can be two very different regimes. If the X-ray luminosity of the central source is high enough ($\sim 10^{37}$~erg/s), the Compton cooling of plasma in the region of the bow shock is rapid, so the matter freely (supersonically) falls toward the magnetosphere, near which the shock is formed. This regime with different degrees of physical description was considered, e.g., in papers by \cite{1976arons}, \cite{1983burnard}, \cite{1991bisno}, or \cite{GX90Kom}. In contrast, the cooling can be ineffective for moderate and low X-ray luminosities, and the matter moves toward the neutron star magnetosphere subsonically, by forming a hot quasi-static shell around the magnetosphere (the settling accretion). Such shells were considered by \cite{1980davies}. A superadiabatic temperature gradient can be established in the shell, leading to development of large-scale convective motions and turbulent cascades. Recently, this regime of accretion has been studied by \cite{shakura2011} \citep[see also first results in][]{postnov2010}. In the settling accretion regime, the accretion rate onto the neutron star is determined by the ability of plasma to enter the magnetosphere via instabilities, and this determines the mean radial velocity of matter in the shell by the mass conservation law. In the last paper it was found that the critical X-ray luminosity below which the settling accretion regime sets in is about $4\times 10^{36}$~erg/s. 
 
In the free-fall accretion regime, the neutron star spin behaviour is determined by the sign of the specific angular momentum of captured matter (prograde or retrograde). Nevertheless, in the case of GX 1+4, this regime seems to be unlikely for two reasons: (a) the X-ray luminosity of the source at the spin-down stage is quite low and (b) the long-term spin-down with short episodic spin-ups are difficult to reconcile with expected features of wind accretion -- although alternation of the prograde/retrograde angular momentum of the captured matter in inhomogeneous stellar wind is possible, the predominance of the retrograde sign is very enigmatic.

In contrast, in the settling accretion regime, the hot convective shell can mediate the angular momentum transfer to/from the neutron star magnetosphere, and the neutron star can spin up or down depending on the sign of the difference of angular velocity of matter near the magnetospheric boundary and that of the magnetosphere itself.  The gas-dynamic problem of spherically-symmetric accretion flow with cooling and heating due to turbulence (generally, anisotropic) was considered in Shakura et al. (2011). It was found that in this regime with increasing X-ray luminosity, the neutron star spin-down can change for spin-up (even abruptly, if $L_x$ exceeds the critical value for the existence of the shell) and vice versa, and the fluctuations in the spin frequency can anti-correlate with flux fluctuations, which indeed follows from the analysis of BATSE and Fermi/GBM observations of GX 1+4 (see above).

It therefore seems very likely that the settling accretion is under way at the present low-luminosity spin-down state of GX 1+4. Then the source must have had higher X-ray luminosities in the preceding long-term spin-up state. Indeed, at high X-ray luminosities, the Compton cooling near the magnetosphere is very strong and a free-fall gap in the matter flow appears above the magnetosphere, the casual connection between the magnetosphere and the shell is lost, and only spin-up of neutron star is possible.



The fluxes reported during the long-term spin-up phase prior to the 1980s and the luminosities derived from these are generally a few times higher than those reported during the still ongoing spin-down phase. Accounting for the differences in instrumentation and energy ranges and the corresponding uncertainties, we arrive at luminosities $L_{\scriptstyle {2-60~keV}}\sim 1-5\times 10^{37}$~erg~s$^{-1}$ for the early data \citep[e.g.,][]{GX81Doty}. After the torque reversal, luminosities in this energy interval mostly remain in the range of a few times $10^{36}$~erg~s$^{-1}$. A flare seen by BATSE \citep{GX97Chack} had a pulsed flux corresponding to  $L_{\scriptstyle {2-60~keV}}\sim 9\times 10^{36}$ ~erg~s$^{-1}$, which indicates a comparable brightness to the early data. We refer to Appendix~\ref{spinluminosity} for a more detailed discussion.

The middle panel of Fig.~\ref{PetersFigure} shows long-term quasi-periodic frequency fluctuations. Their fractional amplitude $\Delta\omega/|\omega_\mathrm{sd}| = (\delta \dot \omega \Delta T) \simeq \pm 0.1$ over a time interval of about one orbital period ($\Delta t\simeq 1.5\times 10^8$~s) corresponds to the fractional derivative ratio $\delta \dot\omega/|\dot\omega_\mathrm{sd}|\simeq 0.3$. Apparently, they are marginally correlated with periastron passages of the binary system, but not clearly correlated with the X-ray flux variations indicated in the bottom panel of Fig.~\ref{PetersFigure}. These long-term fluctuations can be due to smooth variations of the wind density $\rho_\mathrm{w}$ and velocity near the gravitational capture radius.

The short-term spin-up episodes sometimes observed on top of a steady spin-down behaviour \citep[see Fig. 2 in][and Fig.~\ref{GX_BATSE_old} of this paper]{GX97Chack} are correlated with an enhancement of the X-ray flux, in contrast to the negative frequency-flux correlation at the spin-down discussed above (see Figs.~\ref{GXFermi}~and~\ref{CC}). During these short spin-ups, $\dot\omega$ was about half of the average $\dot\omega_\mathrm{su}$ observed during the steady spin-up state of GX 1+4. The X-ray luminosity during these episodic spin-ups was approximately five times higher than the mean X-ray luminosity during the steady spin-down. These facts are consistent with the quasi-spherical accretion model that predicts transitions from spin-down to spin-up with increasing mean accretion rate  and the reversal of flux-period correlation properties (see Shakura et al. 2011 for more detail). However, an increase in the accretion rate by more than one order of magnitude in GX 1+4 could destroy the shell due to rapid radiation cooling and the establishing of the free-fall accretion regime onto the magnetosphere. In that case only spin-up of neutron star is possible.  


\section{Conclusions}
\label{discussions}

We present the results of observing the pulse period behaviour of the symbiotic X-ray binary GX~1+4. New measurements by BeppoSAX, INTEGRAL, and Fermi confirm the continuing overall spin-down (about $\sim 3.4$~s/y) of the neutron star since the torque reversal in the early 1980s. The pulse period has increased by about $\sim$50\% over the past three decades and has reached the highest value ever observed for this source. The X-ray luminosity during the extended spin-down phase has in general been significantly lower than during the spin-up phase of the 1970s (see Appendix~\ref{spinluminosity}).

The global spin-down follows a linear trend in angular frequency with deviations $<2$\% (see Fig.~\ref{PetersFigure}). During this time interval, there have only been brief instances of spin-up observed by BATSE \citep{GX97Chack} related to bright flares where the X-ray luminosity has almost reached the value it had in the 1970s, and a possible spin-up in 2004 observed by INTEGRAL \citep{GX07Ferrigno}.

On top of the long-term spin-up and spin-down trends going on for decades or more independently of the X-ray luminosity, there are pulse period fluctuations on shorter timescales that do show anti-correlation with the source X-ray flux as demonstrated in Section~\ref{Results} and, apparently, marginally correlated with the orbital phase.

As shown in Secs~\ref{accretionmodels} and~\ref{quasisphe}, these observational facts are difficult to reconcile with prograde or retrograde disc accretion but can be explained by assuming quasi-spherical accretion onto the neutron star from the stellar wind of the M-type giant companion.


\begin{Acknowledgments}

Partially based on observations with INTEGRAL, an ESA project with instruments and a science data centre funded by ESA member states (especially the PI countries: Denmark, France, Germany, Italy, Switzerland, Spain), Czech Republic, and Poland, and with the participation of Russia and the USA. We acknowledge support from the Faculty of the European Space Astronomy Centre (ESAC). The Swift/BAT transient monitor results are provided by the Swift/BAT team. We thank Deepto Chakrabarty for providing most of the historical pulse measurements and Jean in't Zand for providing the BeppoSAX/WFC light curves. The work of KP and AK was partially supported by RFBR grant 10-02-00599. M.F. acknowledges partial support from NASA grants NNX08AW06G and NNX11AE24G. The work of AG has been supported by the Spanish MICINN under FPI Fellowship BES-2009-014217 associated to grant AYA2008-06166-C03-03, and partially funded by grants AYA2010-21697-C05-05 and CSD2006-00070 of the Spanish MICINN.

\end{Acknowledgments}

\bibliographystyle{aa}
\bibliography{bibliography}

\pagebreak

\onecolumn

\appendix

\section{Historical spin period and X-ray flux evolution of GX~1+4}
\label{spinluminosity}

In this section we attempt to summarize the evolution of the spin period, as well as of the brightness of GX~1+4. For a table of all pulse period measurements, we refer to Appendix B. The former is rather straightforward, the latter much less so. The difficulty arises for various reasons. First of all, the flux values come from a variety of instruments with different energy ranges and spectral responses. Second, few publications give enough detail to determine the exact spectral shape used in the analysis. Finally, different broad-band observations \citep[e.g.,][]{GX05Naik} do demonstrate clear spectral variation, including a variable absorption.

To get a basis for comparison of flux we used the detailed spectral shapes published by \citet{GX05Naik} and \citet{GX07Ferrigno} to simulate a range of spectra under different conditions. We varied crucial spectral parameters within the uncertainties given for these spectra to cover more of the possible parameter range. From these simulated spectra we then derived simulated fluxes in various energy ranges and used the ratio of these fluxes to derive flux conversion factors between the bands, together with the corresponding uncertainties. Based on this analysis, we converted selected fluxes to the 2--60~keV band, for which fluxes were reported by \citet{GX04Cui} and \citet{GX07Ferrigno} and very close to the SAS~3 band (1.2--55~keV) for which fluxes are reported in \citet{GX81Doty}. 

The uncertainty in such conversions can be fairly high, especially
when going from a narrow band to a broader band. In our analysis
we estimate these to be up to, e.g, $\sim$55\% scaling a 2--6\,keV
flux to the 2--60\,keV band, while allowing for uncertainties in the
soft band spectral shape and the level of absorption at any given time.
Therefore, we concentrate as much as possible on fluxes that can be 
compared directly for our discussion.

For an earlier overview of the flux evolution of GX~1+4 between 1970
and 1988 we refer to Table 2 of \citet{GX89Mc}.

\subsection{Early observation -- the spin-up phase}

After its detection in the 1970s GX~1+4 was spinning up with the fastest rate ($\dot{P}\sim -8.5 \times 10^{-8}$~s\,s$^{-1}$ or $-2.55$~s\,yr$^{-1}$) among the known X-ray pulsars at the time \citep[e.g.,][]{GX81Doty,GX85Elsner,nagase89}. The OSO-7 scans between October 1971 and May 1973 found flux values
in the Argon counters varying between $\sim2$ and $\sim12$ cts\,s$^{-1}$ \citep{OSO7Markert79},
which roughly scales\footnote{See also \texttt{http://www.astronomycafe.net/qadir/q1723.html}} to a flux in the 2--60 keV band of $(3\pm1\,-\,18\pm6)\times 10^{-9}$~ergs~cm$^{-2}$~s$^{-1}$. From SAS3 observations in 1975 and 1976, \citet{GX81Doty} found fluxes in the 1.2--55~keV range of $(4.5\pm1\,-\,22\pm10)\times 10^{-9}$~ergs~cm$^{-2}$~s$^{-1}$, comparable, within the uncertainties, to OSO-8 observations \citep{GX76Becker}.


\subsection{Low state and torque reversal}

EXOSAT failed to detect the source in 1983 and 1984, revealing an extended low state with upper limits on the 2--10~keV flux of $<10^{-10}$~ergs~cm$^{-2}$~s$^{-1}$ \citep{GX83IAU} and $<10^{-11}$~ergs~cm$^{-2}$~s$^{-1}$ \citep{GX88Mukai}, respectively. Even allowing for a factor 3--7 when scaling this to the broader energy range above and accounting for various uncertainties, the source was fainter by two orders of magnitude or more during the spin-up phase.

A first detection by Ginga in March 1987 \citep{GX88Mak,GX89Dotani} found the source still very faint at $\sim10^{-10}$~ergs~cm$^{-2}$~s$^{-1}$ in the 2--20~keV band (scaling to roughly twice this value in 2--60~keV). The pulse period of $110.223\pm0.003$~s was longer than the last previous measurement \citep{GX82Ric}, demonstrating that a torque reversal had taken place and the source was spinning down (see Fig.~\ref{GXhistdata} and Table~\ref{GXPulseHistory}). In addition, the pulse profile of that early observation was peculiar and very different from previously observed ones.

\subsection{Early spin-down phase}

HEXE and Ginga observations between 1987 and 1989 \citep{GX89Dotani,GX90Sakao,GX91Mony} revealed a remarkably constant spin-down rate at $\dot{P}\sim4.5\times 10^{-8}$~s\,s$^{-1}$ in spite of a further overall increase in the X-ray brightness. Ginga fluxes from March 1988 and August 1989 were reported at five to six times that of the first detection in March 1987 \citep{GX90Sakao}. No correlation between spin torque and X-ray flux was reported during this period.

GX~1+4 was observed by GRANAT/SIGMA during 1990 and 1991 \citep{GX91Denis}. The spin-down of the pulsar was confirmed, but at a lower rate ($\dot{P}\sim2.6\times10^{-8}$~s\,s$^{-1}$). The linear spin-down trend was recovered around September 1991 with a slightly increased rate over the one observed before 1990 \citep{GX94Mandrou}. 

Ginga observations in September 1990 and September 1991 showed a decrease in the 2--20~keV X-ray fluxes to $\sim 6$ and $\sim2\times 10^{-10}$~ergs~cm$^{-2}$~s$^{-1}$, respectively, suggesting another low state, together with a drastic increase in $N_{\rm H}$ \citep{GX99Kotani}. Despite this, the spin-down rate was found to be very stable at $\dot{P}\sim (3-4)\times10^{-8}$~s\,s$^{-1}$.

\subsection{Regular monitoring -- spin-down with interruptions}

BATSE onboard CGRO monitored GX~1+4 daily between April 1991 and September 1994. The average X-ray flux in the 20--60~keV range was $\sim2\times 10^{-10}$~ergs~cm$^{-2}$~s$^{-1}$, but interrupted by intermittent bright flares of about 20 day duration \citep{GX97Chack}. During this time a negative correlation between spin-up torque and X-ray flux, i.e., stronger spin-down for higher flux, was observed \citep[e.g.,][]{GX97Paul,GX97Chack,Nelson}. 

In September 1994 an ASCA observation found the source brighter again with a 2--20~keV flux of $\sim9\times 10^{-10}$~ergs~cm$^{-2}$~s$^{-1}$ \citep{GX99Kotani}. One month later, BATSE detected a $\sim$200 day bright state, when the pulsar started to spin-up, resuming the spin-down when the low state was recovered \citep{GX97Chack}.

BATSE observations between May 1995 and June 1996 showed an erratically varying X-ray luminosity, although the pulsar continued to spin-down at a relatively steady rate with a weak negative peak at zero lag in the cross-correlation of torque and flux histories \citep{GX97Chack}. After a period of low activity for about a month, GX~1+4 reappeared briefly in a bright flare, during which it transited to spin-up \citep{GX97Chack}. The \emph{pulsed flux} in this flare was $\sim2\times 10^{-9}$~ergs~cm$^{-2}$~s$^{-1}$ in the 20--60~keV band. According to our analysis, this would correspond to a $\sim(3-4.5)\times 10^{-9}$~ergs~cm$^{-2}$~s$^{-1}$ \emph{pulsed flux} in the 2--60~keV broad band. 

About ten days after this flare, the source dropped below the the detection limit and remained undetected until December 1996. During this low state, a series of RXTE observations also failed to detect GX~1+4, giving an upper limit for the X-ray luminosity that was comparable to the 1983/1984 EXOSAT low state \citep{GX97Chack}.

\citet{GX04Cui} monitored GX~1+4 weekly with RXTE in 2001 and 2002. While the source was ``quite bright'' at the beginning of the year 2002 with fluxes (2--60 keV) of $\sim (2-3)\times 10^{-9}$~ergs~cm$^{-2}$~s$^{-1}$, it made a transition to a faint state in mid June, reaching a minimum flux of $\sim3\times 10^{-11}$~ergs~cm$^{-2}$~s$^{-1}$, again comparable to the EXOSAT upper limits during the time of the torque reversal. At fluxes below $\sim 2\times 10^{-10}$~ergs~cm$^{-2}$~s$^{-1}$ some 
observations had no detectable X-ray pulsations, while others presented a clearly pulsed signal. Towards the end of the year, the X-ray luminosity increased again. A joint Chandra \& RXTE observation in August 2002 found a 2--20 keV X-ray flux of $3.8\times 10^{-11}$~ergs~cm$^{-2}$~s$^{-1}$ \citep{GX05Paul}. The average spin-down rate during this period was $\dot{P}\sim4.4 \times 10^{-8}$~s\,s$^{-1}$ \citep{GX04Cui}.

During INTEGRAL observations between March 2003 and September 2004 \citep{GX07Ferrigno} the source increased its flux by about a factor of 5, with erratic variations of about one order of magnitude. For two time periods with broad-band coverage by the INTEGRAL instruments in February/March and September 2004, \citet{GX07Ferrigno} determined the 2--60~fluxes to be $\sim1.7\times 10^{-9}$~ergs~cm$^{-2}$~s$^{-1}$ and $\sim2.3\times 10^{-9}$~ergs~cm$^{-2}$~s$^{-1}$, respectively.

The secular spin-down trend of GX~1+4 at the time was $\dot{P}\sim6.6 \times 10^{-8}$~s/s, with a spin-up of the source observed during the high-luminosity state in September 2004 \citep{GX07Ferrigno}.

For more recent data we refer to Secs~\ref{Observations}~and~\ref{Results}. 

\Online

\longtab{1}{
\section{GX~1+4 pulse period measurements}\label{AppPulseHistory}
For reference we have collected all pulse period measurements presented in Fig.~\ref{GXhistdata} and give them in Table~\ref{GXPulseHistory} below.
\begin{longtable}{lllllll}
\caption{\label{GXPulseHistory} Pulse period measurements of GX~1+4}\\
\hline\hline
MJD& $P_{spin}$(s)& Instrument& MJD& $P_{spin}$(s)& Instrument\\
\hline\hline
\endfirsthead
\caption{Pulse period measurements of GX~1+4 (continued).}\\
\hline\hline
MJD& $P_{spin}$(s)& Instrument& MJD& $P_{spin}$(s)& Instrument\\
\hline
\endhead
\hline
\endfoot
40875.5 & $ 135.0 \pm 4.0 $ & MIT balloon~$^1$ & 41571.5 & $ 129.0 \pm 0.3 $ & Copernicus~$^2$ \\
41578.5 & $ 129.6 \pm 0.6 $ & Copernicus~$^2$ & 41766.5 & $ 131.7 \pm 4.2 $ & Copernicus~$^2$ \\
42139.5 & $ 128.1 \pm 0.3 $ & Rice balloon~$^3$ & 42670.5 & $ 122.46 \pm 0.03 $ & OSO-8~$^4$ \\
42695.2 & $ 122.34 \pm 0.06 $ & SAS-3~$^5$ & 42772.8 & $ 121.367 \pm 0.004 $ & SAS-3~$^5$ \\
42812.1 & $ 120.6589 \pm 0.0003 $ & SAS-3~$^5$ & 42853.5 & $ 120.493 \pm 0.003 $ & OSO-8~$^6$ \\
42909.3 & $ 120.5 \pm 0.5 $ & NRL balloon~$^7$ & 42963.2 & $ 120.19 \pm 0.05 $ & SAS-3~$^5$ \\
43213.5 & $ 118.873 \pm 0.005 $ & OSO-8~$^6$ & 43244.5 & $ 118.715 \pm 0.005 $ & Ariel 5~$^8$ \\
43470.6 & $ 117.45 \pm 0.20 $ & NRL balloon~$^7$ & 43593.5 & $ 116.49 \pm 0.10 $ & OSO-8~$^6$ \\
43772.0 & $ 114.254 \pm 0.011 $ & OSO-8~$^6$ & 43834.5 & $ 113.75 \pm 0.25 $ & MPI/Tuebingen balloon~$^9$ \\
43952.5 & $ 112.68 \pm 0.03 $ & Einstein/MPC~$^{10}$ & 44062.95 & $ 112.076 \pm 0.003 $ & Ariel 6~$^{11}$ \\

44346.95 & $ 109.668 \pm 0.003 $ & Ariel 6~$^{11}$ & 45060.0 & $ 108.8 \pm 0.2 $ & Balloon experiment~$^{12}$ \\
46052.0 & $ 108.1 \pm 2.6 $ & TIFR balloon~$^{13}$$^{,}$~$^{14}$ & 46754.25 & $ 111.5 \pm 1.2 $ & Tasmania balloon~$^{15}$ \\
46885.0 & $ 110.233 \pm 0.003 $ & Ginga~$^{16}$ & 47099.54 & $ 111.03 \pm 0.02 $ & Mir/Kvant/HEXE~$^{17}$ \\
47247.3 & $ 111.59 \pm 0.02 $ & Ginga~$^{18}$ & 47429.61 & $ 112.359 \pm 0.005 $ & Mir/Kvant/HEXE~$^{17}$ \\
47765.47 & $ 113.626 \pm 0.002 $ & Ginga~$^{19}$ & 48126.8 & $ 114.540 \pm 0.062 $ & GRANAT/ART-P~$^{20}$ \\
48171.7 & $ 114.657 \pm 0.014 $ & GRANAT/ART-P~$^{20}$ & 48337.87 & $ 115.06 \pm 0.03 $ & GRANAT/SIGMA~$^{21}$ \\
48347.52 & $ 115.086 \pm 0.005 $ & GRANAT/SIGMA~$^{21}$ & 48368.5 & $ 115.2280 \pm 0.0016 $ & CGRO/BATSE~$^{22}$ \\
48373.5 & $ 115.2617 \pm 0.0022 $ & CGRO/BATSE~$^{22}$ & 48378.5 & $ 115.2932 \pm 0.0025 $ & CGRO/BATSE~$^{22}$ \\
48383.5 & $ 115.3301 \pm 0.0016 $ & CGRO/BATSE~$^{22}$ & 48388.5 & $ 115.3664 \pm 0.0016 $ & CGRO/BATSE~$^{22}$ \\
48393.5 & $ 115.4112 \pm 0.0014 $ & CGRO/BATSE~$^{22}$ & 48398.5 & $ 115.4567 \pm 0.0015 $ & CGRO/BATSE~$^{22}$ \\
48403.5 & $ 115.5001 \pm 0.0013 $ & CGRO/BATSE~$^{22}$ & 48408.5 & $ 115.5352 \pm 0.0013 $ & CGRO/BATSE~$^{22}$ \\
48413.5 & $ 115.5668 \pm 0.0017 $ & CGRO/BATSE~$^{22}$ & 48418.5 & $ 115.5978 \pm 0.0025 $ & CGRO/BATSE~$^{22}$ \\
48423.5 & $ 115.6323 \pm 0.0021 $ & CGRO/BATSE~$^{22}$ & 48428.5 & $ 115.6617 \pm 0.0018 $ & CGRO/BATSE~$^{22}$ \\
48433.5 & $ 115.7029 \pm 0.0021 $ & CGRO/BATSE~$^{22}$ & 48438.5 & $ 115.7378 \pm 0.0017 $ & CGRO/BATSE~$^{22}$ \\
48443.5 & $ 115.7689 \pm 0.0020 $ & CGRO/BATSE~$^{22}$ & 48448.5 & $ 115.8091 \pm 0.0023 $ & CGRO/BATSE~$^{22}$ \\
48453.5 & $ 115.8482 \pm 0.0017 $ & CGRO/BATSE~$^{22}$ & 48458.5 & $ 115.8836 \pm 0.0020 $ & CGRO/BATSE~$^{22}$ \\
48463.5 & $ 115.9148 \pm 0.0025 $ & CGRO/BATSE~$^{22}$ & 48468.5 & $ 115.9493 \pm 0.0018 $ & CGRO/BATSE~$^{22}$ \\
48473.5 & $ 115.9780 \pm 0.0024 $ & CGRO/BATSE~$^{22}$ & 48478.5 & $ 116.0134 \pm 0.0023 $ & CGRO/BATSE~$^{22}$ \\
48483.5 & $ 116.0383 \pm 0.0061 $ & CGRO/BATSE~$^{22}$ & 48488.5 & $ 116.0648 \pm 0.0052 $ & CGRO/BATSE~$^{22}$ \\
48493.5 & $ 116.0820 \pm 0.0038 $ & CGRO/BATSE~$^{22}$ & 48498.5 & $ 116.1130 \pm 0.0031 $ & CGRO/BATSE~$^{22}$ \\
48503.5 & $ 116.1392 \pm 0.0040 $ & CGRO/BATSE~$^{22}$ & 48508.5 & $ 116.1592 \pm 0.0027 $ & CGRO/BATSE~$^{22}$ \\
48513.5 & $ 116.1799 \pm 0.0022 $ & CGRO/BATSE~$^{22}$ & 48533.5 & $ 116.3084 \pm 0.0044 $ & CGRO/BATSE~$^{22}$ \\
48538.5 & $ 116.3213 \pm 0.0028 $ & CGRO/BATSE~$^{22}$ & 48543.5 & $ 116.3713 \pm 0.0014 $ & CGRO/BATSE~$^{22}$ \\
48548.5 & $ 116.4208 \pm 0.0014 $ & CGRO/BATSE~$^{22}$ & 48553.5 & $ 116.4638 \pm 0.0040 $ & CGRO/BATSE~$^{22}$ \\
48558.5 & $ 116.4749 \pm 0.0052 $ & CGRO/BATSE~$^{22}$ & 48563.5 & $ 116.5092 \pm 0.0029 $ & CGRO/BATSE~$^{22}$ \\
48568.5 & $ 116.5421 \pm 0.0034 $ & CGRO/BATSE~$^{22}$ & 48573.5 & $ 116.5774 \pm 0.0014 $ & CGRO/BATSE~$^{22}$ \\
48578.5 & $ 116.6152 \pm 0.0017 $ & CGRO/BATSE~$^{22}$ & 48583.5 & $ 116.6493 \pm 0.0019 $ & CGRO/BATSE~$^{22}$ \\
48588.5 & $ 116.6926 \pm 0.0041 $ & CGRO/BATSE~$^{22}$ & 48593.5 & $ 116.7240 \pm 0.0028 $ & CGRO/BATSE~$^{22}$ \\
48598.5 & $ 116.7618 \pm 0.0025 $ & CGRO/BATSE~$^{22}$ & 48603.5 & $ 116.7910 \pm 0.0019 $ & CGRO/BATSE~$^{22}$ \\
48608.5 & $ 116.8300 \pm 0.0020 $ & CGRO/BATSE~$^{22}$ & 48613.5 & $ 116.8659 \pm 0.0020 $ & CGRO/BATSE~$^{22}$ \\
48618.5 & $ 116.9093 \pm 0.0016 $ & CGRO/BATSE~$^{22}$ & 48623.5 & $ 116.9486 \pm 0.0017 $ & CGRO/BATSE~$^{22}$ \\
48628.5 & $ 116.9918 \pm 0.0014 $ & CGRO/BATSE~$^{22}$ & 48633.5 & $ 117.0330 \pm 0.0012 $ & CGRO/BATSE~$^{22}$ \\
48638.5 & $ 117.0752 \pm 0.0013 $ & CGRO/BATSE~$^{22}$ & 48643.5 & $ 117.1177 \pm 0.0014 $ & CGRO/BATSE~$^{22}$ \\
48648.5 & $ 117.1563 \pm 0.0017 $ & CGRO/BATSE~$^{22}$ & 48653.5 & $ 117.2024 \pm 0.0019 $ & CGRO/BATSE~$^{22}$ \\
48658.5 & $ 117.2400 \pm 0.0020 $ & CGRO/BATSE~$^{22}$ & 48663.5 & $ 117.2894 \pm 0.0014 $ & CGRO/BATSE~$^{22}$ \\
48668.5 & $ 117.3327 \pm 0.0011 $ & CGRO/BATSE~$^{22}$ & 48671.0 & $ 117.3590 \pm 0.0160 $ & GRANAT/SIGMA~$^{23}$ \\
48673.5 & $ 117.3786 \pm 0.0013 $ & CGRO/BATSE~$^{22}$ & 48678.5 & $ 117.4225 \pm 0.0014 $ & CGRO/BATSE~$^{22}$ \\
48683.5 & $ 117.4645 \pm 0.0015 $ & CGRO/BATSE~$^{22}$ & 48688.5 & $ 117.4987 \pm 0.0030 $ & CGRO/BATSE~$^{22}$ \\
48693.5 & $ 117.5280 \pm 0.0053 $ & CGRO/BATSE~$^{22}$ & 48703.5 & $ 117.6059 \pm 0.0027 $ & CGRO/BATSE~$^{22}$ \\
48708.5 & $ 117.6299 \pm 0.0041 $ & CGRO/BATSE~$^{22}$ & 48713.5 & $ 117.6581 \pm 0.0036 $ & CGRO/BATSE~$^{22}$ \\
48718.5 & $ 117.6943 \pm 0.0024 $ & CGRO/BATSE~$^{22}$ & 48723.5 & $ 117.7281 \pm 0.0034 $ & CGRO/BATSE~$^{22}$ \\
48728.5 & $ 117.7519 \pm 0.0040 $ & CGRO/BATSE~$^{22}$ & 48733.5 & $ 117.7791 \pm 0.0041 $ & CGRO/BATSE~$^{22}$ \\
48738.5 & $ 117.8007 \pm 0.0045 $ & CGRO/BATSE~$^{22}$ & 48743.5 & $ 117.8245 \pm 0.0059 $ & CGRO/BATSE~$^{22}$ \\
48748.5 & $ 117.8547 \pm 0.0026 $ & CGRO/BATSE~$^{22}$ & 48758.5 & $ 117.9050 \pm 0.0038 $ & CGRO/BATSE~$^{22}$ \\
48763.5 & $ 117.9317 \pm 0.0029 $ & CGRO/BATSE~$^{22}$ & 48768.5 & $ 117.9654 \pm 0.0020 $ & CGRO/BATSE~$^{22}$ \\
48773.5 & $ 117.9887 \pm 0.0024 $ & CGRO/BATSE~$^{22}$ & 48778.5 & $ 118.0121 \pm 0.0043 $ & CGRO/BATSE~$^{22}$ \\
48788.5 & $ 118.0692 \pm 0.0045 $ & CGRO/BATSE~$^{22}$ & 48793.5 & $ 118.0842 \pm 0.0048 $ & CGRO/BATSE~$^{22}$ \\
48808.5 & $ 118.1588 \pm 0.0053 $ & CGRO/BATSE~$^{22}$ & 48813.5 & $ 118.1920 \pm 0.0067 $ & CGRO/BATSE~$^{22}$ \\
48818.5 & $ 118.2127 \pm 0.0047 $ & CGRO/BATSE~$^{22}$ & 48823.5 & $ 118.2317 \pm 0.0042 $ & CGRO/BATSE~$^{22}$ \\
48828.5 & $ 118.2550 \pm 0.0061 $ & CGRO/BATSE~$^{22}$ & 48833.5 & $ 118.2585 \pm 0.0083 $ & CGRO/BATSE~$^{22}$ \\
48838.5 & $ 118.2925 \pm 0.0050 $ & CGRO/BATSE~$^{22}$ & 48843.5 & $ 118.3115 \pm 0.0042 $ & CGRO/BATSE~$^{22}$ \\
48848.5 & $ 118.3342 \pm 0.0056 $ & CGRO/BATSE~$^{22}$ & 48858.5 & $ 118.3761 \pm 0.0037 $ & CGRO/BATSE~$^{22}$ \\
48863.5 & $ 118.4001 \pm 0.0072 $ & CGRO/BATSE~$^{22}$ & 48868.5 & $ 118.4083 \pm 0.0043 $ & CGRO/BATSE~$^{22}$ \\
48873.5 & $ 118.4485 \pm 0.0025 $ & CGRO/BATSE~$^{22}$ & 48878.5 & $ 118.4775 \pm 0.0050 $ & CGRO/BATSE~$^{22}$ \\
48883.5 & $ 118.5092 \pm 0.0072 $ & CGRO/BATSE~$^{22}$ & 48888.5 & $ 118.5331 \pm 0.0045 $ & CGRO/BATSE~$^{22}$ \\
48893.5 & $ 118.5587 \pm 0.0033 $ & CGRO/BATSE~$^{22}$ & 48898.5 & $ 118.5938 \pm 0.0024 $ & CGRO/BATSE~$^{22}$ \\
48903.5 & $ 118.6269 \pm 0.0021 $ & CGRO/BATSE~$^{22}$ & 48908.5 & $ 118.6541 \pm 0.0024 $ & CGRO/BATSE~$^{22}$ \\
48913.5 & $ 118.6938 \pm 0.0024 $ & CGRO/BATSE~$^{22}$ & 48918.5 & $ 118.7237 \pm 0.0023 $ & CGRO/BATSE~$^{22}$ \\
48923.5 & $ 118.7449 \pm 0.0040 $ & CGRO/BATSE~$^{22}$ & 48928.5 & $ 118.7706 \pm 0.0037 $ & CGRO/BATSE~$^{22}$ \\
48933.5 & $ 118.7982 \pm 0.0022 $ & CGRO/BATSE~$^{22}$ & 48938.5 & $ 118.8347 \pm 0.0020 $ & CGRO/BATSE~$^{22}$ \\
48943.5 & $ 118.8681 \pm 0.0020 $ & CGRO/BATSE~$^{22}$ & 48948.5 & $ 118.9047 \pm 0.0020 $ & CGRO/BATSE~$^{22}$ \\
48953.5 & $ 118.9359 \pm 0.0027 $ & CGRO/BATSE~$^{22}$ & 48958.5 & $ 118.9641 \pm 0.0045 $ & CGRO/BATSE~$^{22}$ \\
48963.5 & $ 119.0017 \pm 0.0026 $ & CGRO/BATSE~$^{22}$ & 48968.5 & $ 119.0269 \pm 0.0038 $ & CGRO/BATSE~$^{22}$ \\
48973.5 & $ 119.0432 \pm 0.0039 $ & CGRO/BATSE~$^{22}$ & 48978.5 & $ 119.0783 \pm 0.0029 $ & CGRO/BATSE~$^{22}$ \\
48983.5 & $ 119.1106 \pm 0.0023 $ & CGRO/BATSE~$^{22}$ & 48988.5 & $ 119.1483 \pm 0.0019 $ & CGRO/BATSE~$^{22}$ \\
48993.5 & $ 119.1940 \pm 0.0020 $ & CGRO/BATSE~$^{22}$ & 48998.5 & $ 119.2373 \pm 0.0019 $ & CGRO/BATSE~$^{22}$ \\
49003.5 & $ 119.2769 \pm 0.0028 $ & CGRO/BATSE~$^{22}$ & 49008.5 & $ 119.3053 \pm 0.0041 $ & CGRO/BATSE~$^{22}$ \\
49013.5 & $ 119.3430 \pm 0.0039 $ & CGRO/BATSE~$^{22}$ & 49018.5 & $ 119.3886 \pm 0.0025 $ & CGRO/BATSE~$^{22}$ \\
49023.5 & $ 119.4221 \pm 0.0017 $ & CGRO/BATSE~$^{22}$ & 49028.5 & $ 119.4549 \pm 0.0019 $ & CGRO/BATSE~$^{22}$ \\
49033.5 & $ 119.4886 \pm 0.0023 $ & CGRO/BATSE~$^{22}$ & 49038.5 & $ 119.5085 \pm 0.0028 $ & CGRO/BATSE~$^{22}$ \\
49043.5 & $ 119.5367 \pm 0.0044 $ & CGRO/BATSE~$^{22}$ & 49048.5 & $ 119.5562 \pm 0.0027 $ & CGRO/BATSE~$^{22}$ \\
49053.5 & $ 119.5777 \pm 0.0044 $ & CGRO/BATSE~$^{22}$ & 49058.5 & $ 119.5973 \pm 0.0041 $ & CGRO/BATSE~$^{22}$ \\
49063.5 & $ 119.6168 \pm 0.0029 $ & CGRO/BATSE~$^{22}$ & 49068.5 & $ 119.6441 \pm 0.0027 $ & CGRO/BATSE~$^{22}$ \\
49073.5 & $ 119.6639 \pm 0.0030 $ & CGRO/BATSE~$^{22}$ & 49078.5 & $ 119.6938 \pm 0.0031 $ & CGRO/BATSE~$^{22}$ \\
49083.5 & $ 119.7111 \pm 0.0047 $ & CGRO/BATSE~$^{22}$ & 49088.5 & $ 119.7352 \pm 0.0044 $ & CGRO/BATSE~$^{22}$ \\
49093.5 & $ 119.7539 \pm 0.0046 $ & CGRO/BATSE~$^{22}$ & 49098.5 & $ 119.7773 \pm 0.0041 $ & CGRO/BATSE~$^{22}$ \\
49103.5 & $ 119.7868 \pm 0.0078 $ & CGRO/BATSE~$^{22}$ & 49108.5 & $ 119.8148 \pm 0.0029 $ & CGRO/BATSE~$^{22}$ \\
49118.5 & $ 119.8550 \pm 0.0048 $ & CGRO/BATSE~$^{22}$ & 49123.5 & $ 119.8667 \pm 0.0059 $ & CGRO/BATSE~$^{22}$ \\
49128.5 & $ 119.8897 \pm 0.0074 $ & CGRO/BATSE~$^{22}$ & 49133.5 & $ 119.9099 \pm 0.0069 $ & CGRO/BATSE~$^{22}$ \\
49138.5 & $ 119.9292 \pm 0.0039 $ & CGRO/BATSE~$^{22}$ & 49148.5 & $ 119.9736 \pm 0.0038 $ & CGRO/BATSE~$^{22}$ \\
49153.5 & $ 119.9993 \pm 0.0055 $ & CGRO/BATSE~$^{22}$ & 49158.5 & $ 120.0206 \pm 0.0035 $ & CGRO/BATSE~$^{22}$ \\
49163.5 & $ 120.0446 \pm 0.0026 $ & CGRO/BATSE~$^{22}$ & 49168.5 & $ 120.0767 \pm 0.0018 $ & CGRO/BATSE~$^{22}$ \\
49173.5 & $ 120.1002 \pm 0.0014 $ & CGRO/BATSE~$^{22}$ & 49178.5 & $ 120.1274 \pm 0.0015 $ & CGRO/BATSE~$^{22}$ \\
49183.5 & $ 120.1575 \pm 0.0017 $ & CGRO/BATSE~$^{22}$ & 49188.5 & $ 120.1903 \pm 0.0018 $ & CGRO/BATSE~$^{22}$ \\
49193.5 & $ 120.2209 \pm 0.0021 $ & CGRO/BATSE~$^{22}$ & 49198.5 & $ 120.2545 \pm 0.0027 $ & CGRO/BATSE~$^{22}$ \\
49203.5 & $ 120.3001 \pm 0.0042 $ & CGRO/BATSE~$^{22}$ & 49208.5 & $ 120.3250 \pm 0.0057 $ & CGRO/BATSE~$^{22}$ \\
49218.5 & $ 120.4089 \pm 0.0028 $ & CGRO/BATSE~$^{22}$ & 49223.5 & $ 120.4351 \pm 0.0026 $ & CGRO/BATSE~$^{22}$ \\
49228.5 & $ 120.4876 \pm 0.0016 $ & CGRO/BATSE~$^{22}$ & 49232.0 & $ 120.5560 \pm 0.0240 $ & GRANAT/SIGMA~$^{23}$ \\
49233.0 & $ 120.5570 \pm 0.0280 $ & GRANAT/SIGMA~$^{23}$ & 49233.5 & $ 120.5371 \pm 0.0014 $ & CGRO/BATSE~$^{22}$ \\
49236.0 & $ 120.5730 \pm 0.0150 $ & GRANAT/SIGMA~$^{23}$ & 49238.0 & $ 120.5870 \pm 0.0150 $ & GRANAT/SIGMA~$^{23}$ \\
49238.5 & $ 120.5885 \pm 0.0016 $ & CGRO/BATSE~$^{22}$ & 49242.0 & $ 120.6440 \pm 0.0150 $ & GRANAT/SIGMA~$^{23}$ \\
49243.5 & $ 120.6464 \pm 0.0014 $ & CGRO/BATSE~$^{22}$ & 49244.0 & $ 120.6680 \pm 0.0200 $ & GRANAT/SIGMA~$^{23}$ \\
49246.0 & $ 120.6750 \pm 0.0080 $ & GRANAT/SIGMA~$^{23}$ & 49248.5 & $ 120.7016 \pm 0.0030 $ & CGRO/BATSE~$^{22}$ \\
49249.0 & $ 120.7080 \pm 0.0160 $ & GRANAT/SIGMA~$^{23}$ & 49253.5 & $ 120.7252 \pm 0.0043 $ & CGRO/BATSE~$^{22}$ \\
49258.5 & $ 120.7591 \pm 0.0047 $ & CGRO/BATSE~$^{22}$ & 49263.5 & $ 120.7738 \pm 0.0035 $ & CGRO/BATSE~$^{22}$ \\
49268.5 & $ 120.8137 \pm 0.0052 $ & CGRO/BATSE~$^{22}$ & 49273.5 & $ 120.8374 \pm 0.0076 $ & CGRO/BATSE~$^{22}$ \\
49288.5 & $ 120.8429 \pm 0.0043 $ & CGRO/BATSE~$^{22}$ & 49293.5 & $ 120.8426 \pm 0.0033 $ & CGRO/BATSE~$^{22}$ \\
49298.5 & $ 120.8701 \pm 0.0042 $ & CGRO/BATSE~$^{22}$ & 49303.5 & $ 120.9023 \pm 0.0045 $ & CGRO/BATSE~$^{22}$ \\
49308.5 & $ 120.9368 \pm 0.0032 $ & CGRO/BATSE~$^{22}$ & 49313.5 & $ 120.9719 \pm 0.0024 $ & CGRO/BATSE~$^{22}$ \\
49318.5 & $ 121.0084 \pm 0.0020 $ & CGRO/BATSE~$^{22}$ & 49323.5 & $ 121.0422 \pm 0.0019 $ & CGRO/BATSE~$^{22}$ \\
49328.5 & $ 121.0770 \pm 0.0021 $ & CGRO/BATSE~$^{22}$ & 49333.5 & $ 121.1123 \pm 0.0024 $ & CGRO/BATSE~$^{22}$ \\
49338.5 & $ 121.1465 \pm 0.0021 $ & CGRO/BATSE~$^{22}$ & 49343.5 & $ 121.1893 \pm 0.0048 $ & CGRO/BATSE~$^{22}$ \\
49363.5 & $ 121.3394 \pm 0.0043 $ & CGRO/BATSE~$^{22}$ & 49368.5 & $ 121.3456 \pm 0.0072 $ & CGRO/BATSE~$^{22}$ \\
49388.5 & $ 121.3486 \pm 0.0048 $ & CGRO/BATSE~$^{22}$ & 49398.5 & $ 121.3689 \pm 0.0061 $ & CGRO/BATSE~$^{22}$ \\
49403.5 & $ 121.3977 \pm 0.0083 $ & CGRO/BATSE~$^{22}$ & 49408.5 & $ 121.4013 \pm 0.0070 $ & CGRO/BATSE~$^{22}$ \\
49413.5 & $ 121.4083 \pm 0.0055 $ & CGRO/BATSE~$^{22}$ & 49418.5 & $ 121.4174 \pm 0.0045 $ & CGRO/BATSE~$^{22}$ \\
49423.5 & $ 121.4414 \pm 0.0030 $ & CGRO/BATSE~$^{22}$ & 49438.5 & $ 121.4471 \pm 0.0079 $ & CGRO/BATSE~$^{22}$ \\
49443.5 & $ 121.4712 \pm 0.0033 $ & CGRO/BATSE~$^{22}$ & 49448.5 & $ 121.4846 \pm 0.0026 $ & CGRO/BATSE~$^{22}$ \\
49453.5 & $ 121.4922 \pm 0.0027 $ & CGRO/BATSE~$^{22}$ & 49458.5 & $ 121.5273 \pm 0.0055 $ & CGRO/BATSE~$^{22}$ \\
49468.5 & $ 121.5630 \pm 0.0073 $ & CGRO/BATSE~$^{22}$ & 49473.5 & $ 121.5816 \pm 0.0038 $ & CGRO/BATSE~$^{22}$ \\
49478.5 & $ 121.6012 \pm 0.0033 $ & CGRO/BATSE~$^{22}$ & 49483.5 & $ 121.6208 \pm 0.0039 $ & CGRO/BATSE~$^{22}$ \\
49488.5 & $ 121.6330 \pm 0.0045 $ & CGRO/BATSE~$^{22}$ & 49493.5 & $ 121.6247 \pm 0.0074 $ & CGRO/BATSE~$^{22}$ \\
49498.5 & $ 121.6371 \pm 0.0041 $ & CGRO/BATSE~$^{22}$ & 49503.5 & $ 121.6693 \pm 0.0062 $ & CGRO/BATSE~$^{22}$ \\
49508.5 & $ 121.6771 \pm 0.0042 $ & CGRO/BATSE~$^{22}$ & 49513.5 & $ 121.7119 \pm 0.0074 $ & CGRO/BATSE~$^{22}$ \\
49518.5 & $ 121.7488 \pm 0.0057 $ & CGRO/BATSE~$^{22}$ & 49523.5 & $ 121.7540 \pm 0.0043 $ & CGRO/BATSE~$^{22}$ \\
49528.5 & $ 121.7890 \pm 0.0038 $ & CGRO/BATSE~$^{22}$ & 49533.5 & $ 121.8093 \pm 0.0038 $ & CGRO/BATSE~$^{22}$ \\
49538.5 & $ 121.8352 \pm 0.0029 $ & CGRO/BATSE~$^{22}$ & 49548.5 & $ 121.8701 \pm 0.0030 $ & CGRO/BATSE~$^{22}$ \\
49558.5 & $ 121.9098 \pm 0.0022 $ & CGRO/BATSE~$^{22}$ & 49563.5 & $ 121.9172 \pm 0.0016 $ & CGRO/BATSE~$^{22}$ \\
49573.5 & $ 121.9533 \pm 0.0027 $ & CGRO/BATSE~$^{22}$ & 49578.5 & $ 121.9681 \pm 0.0036 $ & CGRO/BATSE~$^{22}$ \\
49583.5 & $ 121.9783 \pm 0.0029 $ & CGRO/BATSE~$^{22}$ & 49588.5 & $ 121.9839 \pm 0.0031 $ & CGRO/BATSE~$^{22}$ \\
49593.5 & $ 121.9993 \pm 0.0046 $ & CGRO/BATSE~$^{22}$ & 49598.5 & $ 122.0037 \pm 0.0033 $ & CGRO/BATSE~$^{22}$ \\
49603.5 & $ 122.0168 \pm 0.0034 $ & CGRO/BATSE~$^{22}$ & 49608.5 & $ 122.0263 \pm 0.0023 $ & CGRO/BATSE~$^{22}$ \\
49613.5 & $ 122.0425 \pm 0.0025 $ & CGRO/BATSE~$^{22}$ & 49618.5 & $ 122.0517 \pm 0.0019 $ & CGRO/BATSE~$^{22}$ \\
49623.5 & $ 122.0657 \pm 0.0018 $ & CGRO/BATSE~$^{22}$ & 49628.5 & $ 122.0727 \pm 0.0025 $ & CGRO/BATSE~$^{22}$ \\
49633.5 & $ 122.0869 \pm 0.0019 $ & CGRO/BATSE~$^{22}$ & 49638.5 & $ 122.0986 \pm 0.0014 $ & CGRO/BATSE~$^{22}$ \\
49643.5 & $ 122.1063 \pm 0.0014 $ & CGRO/BATSE~$^{22}$ & 49648.5 & $ 122.1128 \pm 0.0014 $ & CGRO/BATSE~$^{22}$ \\
49653.5 & $ 122.1176 \pm 0.0014 $ & CGRO/BATSE~$^{22}$ & 49658.5 & $ 122.1159 \pm 0.0012 $ & CGRO/BATSE~$^{22}$ \\
49663.5 & $ 122.1039 \pm 0.0012 $ & CGRO/BATSE~$^{22}$ & 49668.5 & $ 122.0851 \pm 0.0012 $ & CGRO/BATSE~$^{22}$ \\
49673.5 & $ 122.0689 \pm 0.0012 $ & CGRO/BATSE~$^{22}$ & 49678.5 & $ 122.0477 \pm 0.0011 $ & CGRO/BATSE~$^{22}$ \\
49683.5 & $ 122.0290 \pm 0.0010 $ & CGRO/BATSE~$^{22}$ & 49688.5 & $ 122.0089 \pm 0.0010 $ & CGRO/BATSE~$^{22}$ \\
49693.5 & $ 121.9882 \pm 0.0011 $ & CGRO/BATSE~$^{22}$ & 49698.5 & $ 121.9674 \pm 0.0013 $ & CGRO/BATSE~$^{22}$ \\
49703.5 & $ 121.9506 \pm 0.0012 $ & CGRO/BATSE~$^{22}$ & 49708.5 & $ 121.9345 \pm 0.0012 $ & CGRO/BATSE~$^{22}$ \\
49713.5 & $ 121.9157 \pm 0.0012 $ & CGRO/BATSE~$^{22}$ & 49718.5 & $ 121.8965 \pm 0.0012 $ & CGRO/BATSE~$^{22}$ \\
49723.5 & $ 121.8798 \pm 0.0012 $ & CGRO/BATSE~$^{22}$ & 49728.5 & $ 121.8653 \pm 0.0011 $ & CGRO/BATSE~$^{22}$ \\
49733.5 & $ 121.8524 \pm 0.0010 $ & CGRO/BATSE~$^{22}$ & 49738.5 & $ 121.8382 \pm 0.0009 $ & CGRO/BATSE~$^{22}$ \\
49743.5 & $ 121.8264 \pm 0.0013 $ & CGRO/BATSE~$^{22}$ & 49748.5 & $ 121.8116 \pm 0.0013 $ & CGRO/BATSE~$^{22}$ \\
49753.5 & $ 121.8013 \pm 0.0014 $ & CGRO/BATSE~$^{22}$ & 49758.5 & $ 121.7929 \pm 0.0015 $ & CGRO/BATSE~$^{22}$ \\
49763.5 & $ 121.7853 \pm 0.0015 $ & CGRO/BATSE~$^{22}$ & 49768.5 & $ 121.7783 \pm 0.0011 $ & CGRO/BATSE~$^{22}$ \\
49773.5 & $ 121.7763 \pm 0.0011 $ & CGRO/BATSE~$^{22}$ & 49778.5 & $ 121.7751 \pm 0.0015 $ & CGRO/BATSE~$^{22}$ \\
49783.5 & $ 121.7749 \pm 0.0012 $ & CGRO/BATSE~$^{22}$ & 49788.5 & $ 121.7792 \pm 0.0012 $ & CGRO/BATSE~$^{22}$ \\
49793.5 & $ 121.7886 \pm 0.0013 $ & CGRO/BATSE~$^{22}$ & 49798.5 & $ 121.7998 \pm 0.0017 $ & CGRO/BATSE~$^{22}$ \\
49803.5 & $ 121.8107 \pm 0.0019 $ & CGRO/BATSE~$^{22}$ & 49808.5 & $ 121.8269 \pm 0.0028 $ & CGRO/BATSE~$^{22}$ \\
49813.5 & $ 121.8397 \pm 0.0045 $ & CGRO/BATSE~$^{22}$ & 49818.5 & $ 121.8633 \pm 0.0069 $ & CGRO/BATSE~$^{22}$ \\
49823.5 & $ 121.9408 \pm 0.0066 $ & CGRO/BATSE~$^{22}$ & 49833.5 & $ 122.0374 \pm 0.0024 $ & CGRO/BATSE~$^{22}$ \\
49838.5 & $ 122.1106 \pm 0.0017 $ & CGRO/BATSE~$^{22}$ & 49843.5 & $ 122.1481 \pm 0.0045 $ & CGRO/BATSE~$^{22}$ \\
49848.5 & $ 122.1751 \pm 0.0020 $ & CGRO/BATSE~$^{22}$ & 49853.5 & $ 122.2054 \pm 0.0019 $ & CGRO/BATSE~$^{22}$ \\
49858.5 & $ 122.2373 \pm 0.0017 $ & CGRO/BATSE~$^{22}$ & 49863.5 & $ 122.2816 \pm 0.0021 $ & CGRO/BATSE~$^{22}$ \\
49868.5 & $ 122.3200 \pm 0.0017 $ & CGRO/BATSE~$^{22}$ & 49873.5 & $ 122.3576 \pm 0.0030 $ & CGRO/BATSE~$^{22}$ \\
49878.5 & $ 122.3916 \pm 0.0029 $ & CGRO/BATSE~$^{22}$ & 49883.5 & $ 122.4229 \pm 0.0017 $ & CGRO/BATSE~$^{22}$ \\
49888.5 & $ 122.4551 \pm 0.0015 $ & CGRO/BATSE~$^{22}$ & 49893.5 & $ 122.4858 \pm 0.0017 $ & CGRO/BATSE~$^{22}$ \\
49898.5 & $ 122.5197 \pm 0.0021 $ & CGRO/BATSE~$^{22}$ & 49903.5 & $ 122.5389 \pm 0.0022 $ & CGRO/BATSE~$^{22}$ \\
49908.5 & $ 122.5647 \pm 0.0016 $ & CGRO/BATSE~$^{22}$ & 49913.5 & $ 122.5870 \pm 0.0017 $ & CGRO/BATSE~$^{22}$ \\
49918.5 & $ 122.6001 \pm 0.0018 $ & CGRO/BATSE~$^{22}$ & 49923.5 & $ 122.6279 \pm 0.0016 $ & CGRO/BATSE~$^{22}$ \\
49928.5 & $ 122.6701 \pm 0.0019 $ & CGRO/BATSE~$^{22}$ & 49933.5 & $ 122.6951 \pm 0.0014 $ & CGRO/BATSE~$^{22}$ \\
49938.5 & $ 122.7172 \pm 0.0016 $ & CGRO/BATSE~$^{22}$ & 49943.5 & $ 122.7428 \pm 0.0022 $ & CGRO/BATSE~$^{22}$ \\
49948.5 & $ 122.7698 \pm 0.0036 $ & CGRO/BATSE~$^{22}$ & 49953.5 & $ 122.7684 \pm 0.0015 $ & CGRO/BATSE~$^{22}$ \\
49958.5 & $ 122.7916 \pm 0.0016 $ & CGRO/BATSE~$^{22}$ & 49963.5 & $ 122.8126 \pm 0.0014 $ & CGRO/BATSE~$^{22}$ \\
49968.5 & $ 122.8554 \pm 0.0019 $ & CGRO/BATSE~$^{22}$ & 49971.0 & $ 122.8770 \pm 0.0210 $ & GRANAT/SIGMA~$^{23}$ \\
49973.5 & $ 122.8844 \pm 0.0021 $ & CGRO/BATSE~$^{22}$ & 49978.0 & $ 122.9230 \pm 0.0090 $ & GRANAT/SIGMA~$^{23}$ \\
49978.5 & $ 122.9032 \pm 0.0013 $ & CGRO/BATSE~$^{22}$ & 49979.0 & $ 122.9420 \pm 0.0400 $ & GRANAT/SIGMA~$^{23}$ \\
49983.5 & $ 122.9475 \pm 0.0015 $ & CGRO/BATSE~$^{22}$ & 49988.5 & $ 122.9744 \pm 0.0017 $ & CGRO/BATSE~$^{22}$ \\
49993.5 & $ 122.9989 \pm 0.0016 $ & CGRO/BATSE~$^{22}$ & 49998.5 & $ 123.0159 \pm 0.0015 $ & CGRO/BATSE~$^{22}$ \\
50003.5 & $ 123.0368 \pm 0.0013 $ & CGRO/BATSE~$^{22}$ & 50008.5 & $ 123.0714 \pm 0.0017 $ & CGRO/BATSE~$^{22}$ \\
50013.5 & $ 123.1171 \pm 0.0015 $ & CGRO/BATSE~$^{22}$ & 50018.5 & $ 123.1516 \pm 0.0017 $ & CGRO/BATSE~$^{22}$ \\
50023.5 & $ 123.1751 \pm 0.0015 $ & CGRO/BATSE~$^{22}$ & 50028.5 & $ 123.2126 \pm 0.0016 $ & CGRO/BATSE~$^{22}$ \\
50033.5 & $ 123.2422 \pm 0.0017 $ & CGRO/BATSE~$^{22}$ & 50038.5 & $ 123.2704 \pm 0.0016 $ & CGRO/BATSE~$^{22}$ \\
50043.5 & $ 123.2900 \pm 0.0013 $ & CGRO/BATSE~$^{22}$ & 50048.5 & $ 123.3110 \pm 0.0013 $ & CGRO/BATSE~$^{22}$ \\
50053.5 & $ 123.3342 \pm 0.0017 $ & CGRO/BATSE~$^{22}$ & 50058.5 & $ 123.3476 \pm 0.0014 $ & CGRO/BATSE~$^{22}$ \\
50063.5 & $ 123.3717 \pm 0.0015 $ & CGRO/BATSE~$^{22}$ & 50068.5 & $ 123.3866 \pm 0.0018 $ & CGRO/BATSE~$^{22}$ \\
50073.5 & $ 123.4039 \pm 0.0013 $ & CGRO/BATSE~$^{22}$ & 50078.5 & $ 123.4371 \pm 0.0018 $ & CGRO/BATSE~$^{22}$ \\
50083.5 & $ 123.4610 \pm 0.0023 $ & CGRO/BATSE~$^{22}$ & 50088.5 & $ 123.4780 \pm 0.0043 $ & CGRO/BATSE~$^{22}$ \\
50093.5 & $ 123.4832 \pm 0.0013 $ & CGRO/BATSE~$^{22}$ & 50098.5 & $ 123.5011 \pm 0.0016 $ & CGRO/BATSE~$^{22}$ \\
50103.5 & $ 123.5105 \pm 0.0022 $ & CGRO/BATSE~$^{22}$ & 50108.5 & $ 123.5284 \pm 0.0019 $ & CGRO/BATSE~$^{22}$ \\
50113.5 & $ 123.5490 \pm 0.0014 $ & CGRO/BATSE~$^{22}$ & 50118.5 & $ 123.5799 \pm 0.0015 $ & CGRO/BATSE~$^{22}$ \\
50123.5 & $ 123.6067 \pm 0.0018 $ & CGRO/BATSE~$^{22}$ & 50128.5 & $ 123.6319 \pm 0.0014 $ & CGRO/BATSE~$^{22}$ \\
50133.5 & $ 123.6635 \pm 0.0015 $ & CGRO/BATSE~$^{22}$ & 50138.5 & $ 123.6869 \pm 0.0014 $ & CGRO/BATSE~$^{22}$ \\
50143.5 & $ 123.7167 \pm 0.0013 $ & CGRO/BATSE~$^{22}$ & 50148.5 & $ 123.7418 \pm 0.0014 $ & CGRO/BATSE~$^{22}$ \\
50153.5 & $ 123.7545 \pm 0.0012 $ & CGRO/BATSE~$^{22}$ & 50158.5 & $ 123.7712 \pm 0.0013 $ & CGRO/BATSE~$^{22}$ \\
50162.0 & $ 123.7780 \pm 0.0120 $ & GRANAT/SIGMA~$^{23}$ & 50163.5 & $ 123.8032 \pm 0.0013 $ & CGRO/BATSE~$^{22}$ \\
50164.0 & $ 123.8020 \pm 0.0150 $ & GRANAT/SIGMA~$^{23}$ & 50167.0 & $ 123.8060 \pm 0.0160 $ & GRANAT/SIGMA~$^{23}$ \\
50168.0 & $ 123.8250 \pm 0.0200 $ & GRANAT/SIGMA~$^{23}$ & 50168.5 & $ 123.8367 \pm 0.0015 $ & CGRO/BATSE~$^{22}$ \\
50172.0 & $ 123.8590 \pm 0.1300 $ & GRANAT/SIGMA~$^{23}$ & 50173.5 & $ 123.8581 \pm 0.0014 $ & CGRO/BATSE~$^{22}$ \\
50178.5 & $ 123.8745 \pm 0.0015 $ & CGRO/BATSE~$^{22}$ & 50183.5 & $ 123.8930 \pm 0.0017 $ & CGRO/BATSE~$^{22}$ \\
50188.5 & $ 123.9078 \pm 0.0014 $ & CGRO/BATSE~$^{22}$ & 50193.5 & $ 123.9222 \pm 0.0017 $ & CGRO/BATSE~$^{22}$ \\
50198.5 & $ 123.9417 \pm 0.0012 $ & CGRO/BATSE~$^{22}$ & 50203.5 & $ 123.9589 \pm 0.0015 $ & CGRO/BATSE~$^{22}$ \\
50208.5 & $ 123.9910 \pm 0.0013 $ & CGRO/BATSE~$^{22}$ & 50213.5 & $ 124.0170 \pm 0.0019 $ & CGRO/BATSE~$^{22}$ \\
50218.5 & $ 124.0544 \pm 0.0012 $ & CGRO/BATSE~$^{22}$ & 50223.5 & $ 124.0866 \pm 0.0014 $ & CGRO/BATSE~$^{22}$ \\
50228.5 & $ 124.1055 \pm 0.0013 $ & CGRO/BATSE~$^{22}$ & 50233.5 & $ 124.1264 \pm 0.0011 $ & CGRO/BATSE~$^{22}$ \\
50238.5 & $ 124.1540 \pm 0.0022 $ & CGRO/BATSE~$^{22}$ & 50243.5 & $ 124.1762 \pm 0.0022 $ & CGRO/BATSE~$^{22}$ \\
50248.5 & $ 124.1898 \pm 0.0039 $ & CGRO/BATSE~$^{22}$ & 50288.5 & $ 124.3753 \pm 0.0052 $ & CGRO/BATSE~$^{22}$ \\
50293.5 & $ 124.3888 \pm 0.0030 $ & CGRO/BATSE~$^{22}$ & 50298.5 & $ 124.3989 \pm 0.0011 $ & CGRO/BATSE~$^{22}$ \\
50303.5 & $ 124.3883 \pm 0.0009 $ & CGRO/BATSE~$^{22}$ & 50308.5 & $ 124.3946 \pm 0.0011 $ & CGRO/BATSE~$^{22}$ \\
50313.0 & $ 124.4040 \pm 0.0030 $ & BeppoSAX~$^{24}$ & 50313.5 & $ 124.4070 \pm 0.0015 $ & CGRO/BATSE~$^{22}$ \\
50418.5 & $ 124.9810 \pm 0.0057 $ & CGRO/BATSE~$^{22}$ & 50423.5 & $ 125.0238 \pm 0.0035 $ & CGRO/BATSE~$^{22}$ \\
50428.5 & $ 125.0702 \pm 0.0031 $ & CGRO/BATSE~$^{22}$ & 50438.5 & $ 125.1725 \pm 0.0033 $ & CGRO/BATSE~$^{22}$ \\
50448.5 & $ 125.2853 \pm 0.0030 $ & CGRO/BATSE~$^{22}$ & 50453.5 & $ 125.3332 \pm 0.0021 $ & CGRO/BATSE~$^{22}$ \\
50458.5 & $ 125.3840 \pm 0.0019 $ & CGRO/BATSE~$^{22}$ & 50463.5 & $ 125.4389 \pm 0.0025 $ & CGRO/BATSE~$^{22}$ \\
      50467.9 & $  125.4870 \pm     0.0022$ &CGRO/BATSE~$^{29}$  &      50475.9 & $  125.5610 \pm     0.0024$ &CGRO/BATSE~$^{29}$  \\
      50519.9 & $  125.9130 \pm     0.0024$ &CGRO/BATSE~$^{29}$  &	50532.0 & $ 126.0180 \pm 0.0080 $ & BeppoSAX/LECs,MECs,PDS~$^{24}$\\
      50536.5 & $  126.0380 \pm     0.0021$ &CGRO/BATSE~$^{29}$  &        50539.7 & $  126.0660 \pm     0.0021$ &CGRO/BATSE~$^{29}$  \\ 
      50544.0 & $  126.1150 \pm     0.0020$ &CGRO/BATSE~$^{29}$  &        50567.9 & $  126.2750 \pm     0.0020$ &CGRO/BATSE~$^{29}$  \\ 
      50572.0 & $  126.3020 \pm     0.0030$ &CGRO/BATSE~$^{29}$  &        50580.0 & $  126.3730 \pm     0.0030$ &CGRO/BATSE~$^{29}$  \\ 
      50584.1 & $  126.4040 \pm     0.0009$ &CGRO/BATSE~$^{29}$  &        50588.0 & $  126.4540 \pm     0.0009$ &CGRO/BATSE~$^{29}$  \\ 
      50592.1 & $  126.5140 \pm     0.0007$ &CGRO/BATSE~$^{29}$  &        50595.6 & $  126.5630 \pm     0.0007$ &CGRO/BATSE~$^{29}$  \\ 
      50599.9 & $  126.6010 \pm     0.0020$ &CGRO/BATSE~$^{29}$  &        50604.3 & $  126.6410 \pm     0.0020$ &CGRO/BATSE~$^{29}$  \\ 
      50607.9 & $  126.6690 \pm     0.0021$ &CGRO/BATSE~$^{29}$  &        50616.0 & $  126.7560 \pm     0.0021$ &CGRO/BATSE~$^{29}$  \\ 
      50620.0 & $  126.8060 \pm     0.0012$ &CGRO/BATSE~$^{29}$  &        50624.1 & $  126.8470 \pm     0.0012$ &CGRO/BATSE~$^{29}$  \\ 
      50628.0 & $  126.8850 \pm     0.0022$ &CGRO/BATSE~$^{29}$  &        50632.0 & $  126.9350 \pm     0.0022$ &CGRO/BATSE~$^{29}$  \\ 
      50636.0 & $  126.9590 \pm     0.0006$ &CGRO/BATSE~$^{29}$  &        50640.0 & $  126.9740 \pm     0.0006$ &CGRO/BATSE~$^{29}$  \\ 
      50643.8 & $  126.9800 \pm     0.0007$ &CGRO/BATSE~$^{29}$  &        50648.0 & $  127.0000 \pm     0.0007$ &CGRO/BATSE~$^{29}$  \\ 
      50652.0 & $  127.0170 \pm     0.0006$ &CGRO/BATSE~$^{29}$  &        50656.0 & $  127.0390 \pm     0.0006$ &CGRO/BATSE~$^{29}$  \\ 
      50660.0 & $  127.0550 \pm     0.0005$ &CGRO/BATSE~$^{29}$  &        50663.9 & $  127.0650 \pm     0.0005$ &CGRO/BATSE~$^{29}$  \\ 
      50668.0 & $  127.0850 \pm     0.0007$ &CGRO/BATSE~$^{29}$  &        50672.2 & $  127.0990 \pm     0.0007$ &CGRO/BATSE~$^{29}$  \\ 
      50676.0 & $  127.1150 \pm     0.0007$ &CGRO/BATSE~$^{29}$  &        50680.0 & $  127.1350 \pm     0.0007$ &CGRO/BATSE~$^{29}$  \\ 
      50684.0 & $  127.1420 \pm     0.0012$ &CGRO/BATSE~$^{29}$  &        50688.0 & $  127.1770 \pm     0.0012$ &CGRO/BATSE~$^{29}$  \\ 
      50692.0 & $  127.1810 \pm     0.0022$ &CGRO/BATSE~$^{29}$  &        50704.0 & $  127.2630 \pm     0.0022$ &CGRO/BATSE~$^{29}$  \\ 
      50708.0 & $  127.2850 \pm     0.0027$ &CGRO/BATSE~$^{29}$  &        50720.0 & $  127.3670 \pm     0.0027$ &CGRO/BATSE~$^{29}$  \\ 
      50724.0 & $  127.3660 \pm     0.0024$ &CGRO/BATSE~$^{29}$  &        50728.0 & $  127.3800 \pm     0.0024$ &CGRO/BATSE~$^{29}$  \\ 
      50732.0 & $  127.4080 \pm     0.0020$ &CGRO/BATSE~$^{29}$  &        50768.0 & $  127.5600 \pm     0.0020$ &CGRO/BATSE~$^{29}$  \\ 
      50772.1 & $  127.5710 \pm     0.0019$ &CGRO/BATSE~$^{29}$  &        50776.0 & $  127.5770 \pm     0.0019$ &CGRO/BATSE~$^{29}$  \\ 
      50780.0 & $  127.5870 \pm     0.0011$ &CGRO/BATSE~$^{29}$  &        50784.0 & $  127.5970 \pm     0.0011$ &CGRO/BATSE~$^{29}$  \\ 
      50788.0 & $  127.6060 \pm     0.0014$ &CGRO/BATSE~$^{29}$  &        50791.9 & $  127.6150 \pm     0.0014$ &CGRO/BATSE~$^{29}$  \\ 
      50795.9 & $  127.6210 \pm     0.0011$ &CGRO/BATSE~$^{29}$  &        50800.0 & $  127.6280 \pm     0.0011$ &CGRO/BATSE~$^{29}$  \\ 
      50804.0 & $  127.6370 \pm     0.0010$ &CGRO/BATSE~$^{29}$  &        50808.0 & $  127.6450 \pm     0.0010$ &CGRO/BATSE~$^{29}$  \\ 
      50812.0 & $  127.6520 \pm     0.0011$ &CGRO/BATSE~$^{29}$  &        50816.1 & $  127.6570 \pm     0.0011$ &CGRO/BATSE~$^{29}$  \\ 
      50820.1 & $  127.6680 \pm     0.0014$ &CGRO/BATSE~$^{29}$  &        50824.1 & $  127.6750 \pm     0.0014$ &CGRO/BATSE~$^{29}$  \\ 
      50828.1 & $  127.6830 \pm     0.0015$ &CGRO/BATSE~$^{29}$  &        50832.0 & $  127.6910 \pm     0.0015$ &CGRO/BATSE~$^{29}$  \\ 
      50835.9 & $  127.6970 \pm     0.0013$ &CGRO/BATSE~$^{29}$  &        50839.9 & $  127.7030 \pm     0.0013$ &CGRO/BATSE~$^{29}$  \\ 
      50844.0 & $  127.7080 \pm     0.0012$ &CGRO/BATSE~$^{29}$  &        50847.9 & $  127.7180 \pm     0.0012$ &CGRO/BATSE~$^{29}$  \\ 
      50852.0 & $  127.7250 \pm     0.0013$ &CGRO/BATSE~$^{29}$  &        50856.0 & $  127.7320 \pm     0.0013$ &CGRO/BATSE~$^{29}$  \\ 
      50860.0 & $  127.7370 \pm     0.0018$ &CGRO/BATSE~$^{29}$  &        50864.1 & $  127.7470 \pm     0.0018$ &CGRO/BATSE~$^{29}$  \\ 
      50868.1 & $  127.7520 \pm     0.0024$ &CGRO/BATSE~$^{29}$  &        50872.0 & $  127.7530 \pm     0.0024$ &CGRO/BATSE~$^{29}$  \\ 
      50876.0 & $  127.7830 \pm     0.0021$ &CGRO/BATSE~$^{29}$  &        50880.1 & $  127.7690 \pm     0.0021$ &CGRO/BATSE~$^{29}$  \\ 
      50884.1 & $  127.7710 \pm     0.0029$ &CGRO/BATSE~$^{29}$  &        50887.9 & $  127.7820 \pm     0.0029$ &CGRO/BATSE~$^{29}$  \\ 
      50900.0 & $  127.7960 \pm     0.0019$ &CGRO/BATSE~$^{29}$  &        50916.0 & $  127.8070 \pm     0.0019$ &CGRO/BATSE~$^{29}$  \\ 
      50920.1 & $  127.8120 \pm     0.0073$ &CGRO/BATSE~$^{29}$  &        50924.0 & $  127.8150 \pm     0.0073$ &CGRO/BATSE~$^{29}$  \\ 
      50928.0 & $  127.8120 \pm     0.0020$ &CGRO/BATSE~$^{29}$  &        50932.0 & $  127.8100 \pm     0.0020$ &CGRO/BATSE~$^{29}$  \\ 
      50935.9 & $  127.8070 \pm     0.0014$ &CGRO/BATSE~$^{29}$  &        50940.0 & $  127.8060 \pm     0.0014$ &CGRO/BATSE~$^{29}$  \\ 
      50944.0 & $  127.7960 \pm     0.0009$ &CGRO/BATSE~$^{29}$  &        50948.0 & $  127.7890 \pm     0.0009$ &CGRO/BATSE~$^{29}$  \\ 
      50951.9 & $  127.7850 \pm     0.0015$ &CGRO/BATSE~$^{29}$  &        50959.9 & $  127.7770 \pm     0.0015$ &CGRO/BATSE~$^{29}$  \\ 
      50964.1 & $  127.8110 \pm     0.0015$ &CGRO/BATSE~$^{29}$  &        50968.0 & $  127.8420 \pm     0.0015$ &CGRO/BATSE~$^{29}$  \\ 
      50972.0 & $  127.8710 \pm     0.0011$ &CGRO/BATSE~$^{29}$  &        50976.0 & $  127.9080 \pm     0.0011$ &CGRO/BATSE~$^{29}$  \\ 
      50980.0 & $  127.9390 \pm     0.0007$ &CGRO/BATSE~$^{29}$  &        50984.0 & $  127.9730 \pm     0.0007$ &CGRO/BATSE~$^{29}$  \\ 
      50988.0 & $  128.0080 \pm     0.0006$ &CGRO/BATSE~$^{29}$  &        50992.1 & $  128.0420 \pm     0.0006$ &CGRO/BATSE~$^{29}$  \\ 
      50996.0 & $  128.0750 \pm     0.0007$ &CGRO/BATSE~$^{29}$  &        50999.5 & $  128.1000 \pm     0.0007$ &CGRO/BATSE~$^{29}$  \\ 
      51004.8 & $  128.1520 \pm     0.0015$ &CGRO/BATSE~$^{29}$  &        51008.0 & $  128.1780 \pm     0.0015$ &CGRO/BATSE~$^{29}$  \\ 
      51012.1 & $  128.2140 \pm     0.0007$ &CGRO/BATSE~$^{29}$  &        51016.0 & $  128.2520 \pm     0.0007$ &CGRO/BATSE~$^{29}$  \\ 
      51020.1 & $  128.2880 \pm     0.0007$ &CGRO/BATSE~$^{29}$  &        51023.9 & $  128.3210 \pm     0.0007$ &CGRO/BATSE~$^{29}$  \\ 
      51028.0 & $  128.3560 \pm     0.0007$ &CGRO/BATSE~$^{29}$  &        51032.0 & $  128.3920 \pm     0.0007$ &CGRO/BATSE~$^{29}$  \\ 
      51036.0 & $  128.4270 \pm     0.0006$ &CGRO/BATSE~$^{29}$  &        51040.0 & $  128.4640 \pm     0.0006$ &CGRO/BATSE~$^{29}$  \\ 
      51044.0 & $  128.5010 \pm     0.0004$ &CGRO/BATSE~$^{29}$  &        51046.6 & $  128.5250 \pm     0.0004$ &CGRO/BATSE~$^{29}$  \\ 
      51047.8 & $ 128.5442 \pm 0.0107 $ & BeppoSAX/WFC~$^{25}$ &	51052.9 & $  128.5840 \pm     0.0008$ &CGRO/BATSE~$^{29}$  \\
      51056.0 & $  128.6140 \pm     0.0008$ &CGRO/BATSE~$^{29}$  & 	51057.8 & $ 128.6338 \pm 0.0053 $ & BeppoSAX/WFC~$^{25}$\\
      51060.1 & $  128.6530 \pm     0.0005$ &CGRO/BATSE~$^{29}$  &        51063.9 & $  128.6920 \pm     0.0005$ &CGRO/BATSE~$^{29}$  \\ 
      51067.2 & $ 128.7238 \pm 0.0049 $ & BeppoSAX/WFC~$^{25}$ &	51068.0 & $  128.7300 \pm     0.0004$ &CGRO/BATSE~$^{29}$  \\
      51072.1 & $  128.7680 \pm     0.0004$ &CGRO/BATSE~$^{29}$  &      51075.9 & $  128.8050 \pm     0.0004$ &CGRO/BATSE~$^{29}$  \\
      51079.8 & $  128.8410 \pm     0.0004$ &CGRO/BATSE~$^{29}$  &    51079.2 & $ 128.8189 \pm 0.0127 $ & BeppoSAX/WFC~$^{25}$  \\
      51084.0 & $  128.8820 \pm     0.0003$ &CGRO/BATSE~$^{29}$  &     51084.6 & $ 128.8474 \pm 0.0240 $ & BeppoSAX/WFC~$^{25}$   \\
      51088.0 & $  128.9190 \pm     0.0003$ &CGRO/BATSE~$^{29}$  &      51092.0 & $  128.9570 \pm     0.0003$ &CGRO/BATSE~$^{29}$  \\
      51092.3 & $ 128.9581 \pm 0.0033 $ & BeppoSAX/WFC~$^{25}$ &	 51096.0 & $  128.9920 \pm     0.0003$ &CGRO/BATSE~$^{29}$  \\ 
      51099.9 & $  129.0240 \pm     0.0002$ &CGRO/BATSE~$^{29}$  &        51103.8 & $  129.0580 \pm     0.0002$ &CGRO/BATSE~$^{29}$  \\ 
      51107.9 & $  129.0890 \pm     0.0006$ &CGRO/BATSE~$^{29}$  &        51112.0 & $  129.1230 \pm     0.0006$ &CGRO/BATSE~$^{29}$  \\ 
      51116.0 & $  129.1520 \pm     0.0008$ &CGRO/BATSE~$^{29}$  &        51120.1 & $  129.1840 \pm     0.0008$ &CGRO/BATSE~$^{29}$  \\ 
      51124.0 & $  129.2150 \pm     0.0010$ &CGRO/BATSE~$^{29}$  &        51128.1 & $  129.2460 \pm     0.0010$ &CGRO/BATSE~$^{29}$  \\ 
      51132.0 & $  129.2780 \pm     0.0008$ &CGRO/BATSE~$^{29}$  &        51136.1 & $  129.3100 \pm     0.0008$ &CGRO/BATSE~$^{29}$  \\ 
      51139.9 & $  129.3420 \pm     0.0008$ &CGRO/BATSE~$^{29}$  &        51144.1 & $  129.3740 \pm     0.0008$ &CGRO/BATSE~$^{29}$  \\ 
      51147.9 & $  129.4040 \pm     0.0008$ &CGRO/BATSE~$^{29}$  &        51152.0 & $  129.4420 \pm     0.0008$ &CGRO/BATSE~$^{29}$  \\ 
      51156.0 & $  129.4770 \pm     0.0005$ &CGRO/BATSE~$^{29}$  &        51160.0 & $  129.5110 \pm     0.0005$ &CGRO/BATSE~$^{29}$  \\ 
      51163.8 & $  129.5450 \pm     0.0009$ &CGRO/BATSE~$^{29}$  &        51168.0 & $  129.5790 \pm     0.0009$ &CGRO/BATSE~$^{29}$  \\ 
      51172.0 & $  129.6100 \pm     0.0008$ &CGRO/BATSE~$^{29}$  &        51176.0 & $  129.6470 \pm     0.0008$ &CGRO/BATSE~$^{29}$  \\ 
      51180.0 & $  129.6830 \pm     0.0008$ &CGRO/BATSE~$^{29}$  &        51184.0 & $  129.7190 \pm     0.0008$ &CGRO/BATSE~$^{29}$  \\ 
      51188.1 & $  129.7520 \pm     0.0007$ &CGRO/BATSE~$^{29}$  &        51192.0 & $  129.7870 \pm     0.0007$ &CGRO/BATSE~$^{29}$  \\ 
      51196.1 & $  129.8230 \pm     0.0008$ &CGRO/BATSE~$^{29}$  &        51200.0 & $  129.8580 \pm     0.0008$ &CGRO/BATSE~$^{29}$  \\ 
      51204.0 & $  129.8940 \pm     0.0005$ &CGRO/BATSE~$^{29}$  &        51208.0 & $  129.9310 \pm     0.0005$ &CGRO/BATSE~$^{29}$  \\ 
      51212.0 & $  129.9650 \pm     0.0005$ &CGRO/BATSE~$^{29}$  &        51216.0 & $  130.0030 \pm     0.0005$ &CGRO/BATSE~$^{29}$  \\ 
      51219.9 & $  130.0370 \pm     0.0005$ &CGRO/BATSE~$^{29}$  &        51224.0 & $  130.0750 \pm     0.0005$ &CGRO/BATSE~$^{29}$  \\ 
      51228.0 & $  130.1120 \pm     0.0005$ &CGRO/BATSE~$^{29}$  &        51230.3 & $ 130.1523 \pm 0.0074 $ & BeppoSAX/WFC~$^{25}$ \\
	51232.0 & $  130.1530 \pm     0.0005$ &CGRO/BATSE~$^{29}$  &      51236.1 & $  130.1930 \pm     0.0005$ &CGRO/BATSE~$^{29}$  \\
      51240.0 & $  130.2320 \pm     0.0005$ &CGRO/BATSE~$^{29}$  &      51244.1 & $  130.2730 \pm     0.0004$ &CGRO/BATSE~$^{29}$  \\
       51248.1 & $  130.3150 \pm     0.0004$ &CGRO/BATSE~$^{29}$  &	 51249.4 & $ 130.3295 \pm 0.0140 $ & BeppoSAX/WFC~$^{25}$ \\
      51251.9 & $  130.3560 \pm     0.0004$ &CGRO/BATSE~$^{29}$  &        51256.0 & $  130.3990 \pm     0.0004$ &CGRO/BATSE~$^{29}$  \\ 
      51259.9 & $  130.4410 \pm     0.0003$ &CGRO/BATSE~$^{29}$  &        51262.5 & $ 130.4689 \pm 0.0131 $ & BeppoSAX/WFC~$^{25}$ \\
	51263.9 & $  130.4880 \pm     0.0003$ &CGRO/BATSE~$^{29}$  &      51267.9 & $  130.5330 \pm     0.0003$ &CGRO/BATSE~$^{29}$  \\
       51270.8 & $ 130.5625 \pm 0.0074 $ & BeppoSAX/WFC~$^{25}$ &	51272.0 & $  130.5780 \pm     0.0003$ &CGRO/BATSE~$^{29}$  \\ 
      51274.7 & $ 130.6058 \pm 0.0099 $ & BeppoSAX/WFC~$^{25}$  &  51276.1 & $  130.6250 \pm     0.0003$ &CGRO/BATSE~$^{29}$ \\
      51278.5 & $ 130.6510 \pm 0.0047 $ & BeppoSAX/WFC~$^{25}$  &	 51280.1 & $  130.6720 \pm     0.0003$ &CGRO/BATSE~$^{29}$  \\ 
      51284.1 & $  130.7200 \pm     0.0003$ &CGRO/BATSE~$^{29}$  &        51288.0 & $  130.7660 \pm     0.0003$ &CGRO/BATSE~$^{29}$  \\ 
      51292.0 & $  130.8130 \pm     0.0002$ &CGRO/BATSE~$^{29}$  &        51296.1 & $  130.8590 \pm     0.0002$ &CGRO/BATSE~$^{29}$  \\ 
      51300.1 & $  130.9050 \pm     0.0002$ &CGRO/BATSE~$^{29}$  &        51304.0 & $  130.9530 \pm     0.0002$ &CGRO/BATSE~$^{29}$  \\ 
      51308.0 & $  130.9990 \pm     0.0002$ &CGRO/BATSE~$^{29}$  &        51312.0 & $  131.0450 \pm     0.0002$ &CGRO/BATSE~$^{29}$  \\ 
      51315.9 & $  131.0910 \pm     0.0003$ &CGRO/BATSE~$^{29}$  &        51320.0 & $  131.1390 \pm     0.0003$ &CGRO/BATSE~$^{29}$  \\ 
      51324.0 & $  131.1850 \pm     0.0003$ &CGRO/BATSE~$^{29}$  &        51328.0 & $  131.2280 \pm     0.0003$ &CGRO/BATSE~$^{29}$  \\ 
      51332.0 & $  131.2740 \pm     0.0003$ &CGRO/BATSE~$^{29}$  &        51336.0 & $  131.3210 \pm     0.0003$ &CGRO/BATSE~$^{29}$  \\ 
      51340.0 & $  131.3650 \pm     0.0004$ &CGRO/BATSE~$^{29}$  &        51344.0 & $  131.4090 \pm     0.0004$ &CGRO/BATSE~$^{29}$  \\ 
51460.9 & $ 132.5723 \pm 0.0089 $ & BeppoSAX/WFC~$^{25}$ & 51464.3 & $ 132.6220 \pm 0.0174 $ & BeppoSAX/WFC~$^{25}$ \\
51785.0 & $ 134.9256 \pm 0.0010 $ & BeppoSAX~$^{24}$ & 52346.1 & $ 137.4690 \pm 0.3680 $ & RXTE/PCA\&HEXTE~$^{26}$ \\
52351.3 & $ 137.2540 \pm 0.2910 $ & RXTE/PCA\&HEXTE~$^{26}$ & 52359.9 & $ 137.7750 \pm 0.0920 $ & RXTE/PCA\&HEXTE~$^{26}$ \\
52366.6 & $ 137.9440 \pm 0.2760 $ & RXTE/PCA\&HEXTE~$^{26}$ & 52384.7 & $ 137.8050 \pm 0.1690 $ & RXTE/PCA\&HEXTE~$^{26}$ \\
52392.3 & $ 137.9580 \pm 0.1840 $ & RXTE/PCA\&HEXTE~$^{26}$ & 52401.2 & $ 137.8650 \pm 0.2000 $ & RXTE/PCA\&HEXTE~$^{26}$ \\
52407.0 & $ 137.6350 \pm 0.1840 $ & RXTE/PCA\&HEXTE~$^{26}$ & 52418.0 & $ 138.0180 \pm 0.2610 $ & RXTE/PCA\&HEXTE~$^{26}$ \\
52422.0 & $ 137.1890 \pm 0.2910 $ & RXTE/PCA\&HEXTE~$^{26}$ & 52431.2 & $ 138.2020 \pm 0.4760 $ & RXTE/PCA\&HEXTE~$^{26}$ \\
52434.9 & $ 138.0790 \pm 0.1990 $ & RXTE/PCA\&HEXTE~$^{26}$ & 52450.8 & $ 137.5860 \pm 1.8740 $ & RXTE/PCA\&HEXTE~$^{26}$ \\
52466.1 & $ 137.1250 \pm 0.3070 $ & RXTE/PCA\&HEXTE~$^{26}$ & 52470.7 & $ 137.9390 \pm 0.1840 $ & RXTE/PCA\&HEXTE~$^{26}$ \\
52490.9 & $ 138.1370 \pm 0.2300 $ & RXTE/PCA\&HEXTE~$^{26}$ & 52491.0 & $ 138.1700 \pm 0.0010 $ & Chandra\&RXTE~$^{27}$ \\
52493.0 & $ 138.1680 \pm 0.1230 $ & RXTE/PCA\&HEXTE~$^{26}$ & 52499.2 & $ 138.1060 \pm 0.1690 $ & RXTE/PCA\&HEXTE~$^{26}$ \\
52502.2 & $ 138.1060 \pm 0.5530 $ & RXTE/PCA\&HEXTE~$^{26}$ & 52503.5 & $ 138.2590 \pm 0.8450 $ & RXTE/PCA\&HEXTE~$^{26}$ \\
52512.6 & $ 138.7040 \pm 0.4150 $ & RXTE/PCA\&HEXTE~$^{26}$ & 52515.7 & $ 138.3350 \pm 0.1850 $ & RXTE/PCA\&HEXTE~$^{26}$ \\
52521.8 & $ 138.3660 \pm 0.4610 $ & RXTE/PCA\&HEXTE~$^{26}$ & 52526.1 & $ 138.3960 \pm 0.4920 $ & RXTE/PCA\&HEXTE~$^{26}$ \\
52528.9 & $ 138.0580 \pm 0.6760 $ & RXTE/PCA\&HEXTE~$^{26}$ & 52535.0 & $ 138.1350 \pm 0.2610 $ & RXTE/PCA\&HEXTE~$^{26}$ \\
52539.6 & $ 138.4570 \pm 0.3840 $ & RXTE/PCA\&HEXTE~$^{26}$ & 52542.3 & $ 138.1650 \pm 0.6450 $ & RXTE/PCA\&HEXTE~$^{26}$ \\
52548.4 & $ 137.9960 \pm 0.5680 $ & RXTE/PCA\&HEXTE~$^{26}$ & 52555.5 & $ 138.0570 \pm 3.2090 $ & RXTE/PCA\&HEXTE~$^{26}$ \\
52561.0 & $ 138.0570 \pm 1.5660 $ & RXTE/PCA\&HEXTE~$^{26}$ & 52565.6 & $ 138.1180 \pm 0.7980 $ & RXTE/PCA\&HEXTE~$^{26}$ \\
52582.1 & $ 138.1630 \pm 0.4920 $ & RXTE/PCA\&HEXTE~$^{26}$ & 52585.8 & $ 138.6700 \pm 0.2910 $ & RXTE/PCA\&HEXTE~$^{26}$ \\
52589.2 & $ 138.6230 \pm 0.1080 $ & RXTE/PCA\&HEXTE~$^{26}$ & 52591.9 & $ 138.5770 \pm 0.2000 $ & RXTE/PCA\&HEXTE~$^{26}$ \\
52594.1 & $ 138.4850 \pm 0.0920 $ & RXTE/PCA\&HEXTE~$^{26}$ & 52917.6 & $ 139.6300 \pm 0.0060 $ & INTEGRAL~$^{28}$ \\
53052.1 & $ 140.6132 \pm 0.0002 $ & INTEGRAL~$^{28}$ & 53252.4 & $ 141.5649 \pm 0.0001 $ & INTEGRAL~$^{28}$ \\
53422.0 & $ 141.3310 \pm 0.3380 $ & INTEGRAL/ISGRI~$^{25}$ & 53431.0 & $ 140.9101 \pm 0.5285 $ & INTEGRAL/ISGRI~$^{25}$ \\
53434.0 & $ 141.3730 \pm 0.2320 $ & INTEGRAL/ISGRI~$^{25}$ & 53464.0 & $ 141.8630 \pm 0.2986 $ & INTEGRAL/ISGRI~$^{25}$ \\
53467.0 & $ 141.6670 \pm 0.1540 $ & INTEGRAL/ISGRI~$^{25}$ & 53470.0 & $ 141.4530 \pm 0.2489 $ & INTEGRAL/ISGRI~$^{25}$ \\
53473.0 & $ 141.5861 \pm 0.3421 $ & INTEGRAL/ISGRI~$^{25}$ & 53476.0 & $ 141.6270 \pm 0.1014 $ & INTEGRAL/ISGRI~$^{25}$ \\
53479.0 & $ 141.8120 \pm 0.1796 $ & INTEGRAL/ISGRI~$^{25}$ & 53611.0 & $ 142.9630 \pm 0.2911 $ & INTEGRAL/ISGRI~$^{25}$ \\
53620.0 & $ 142.9510 \pm 0.3223 $ & INTEGRAL/ISGRI~$^{25}$ & 53635.0 & $ 142.8820 \pm 0.1238 $ & INTEGRAL/ISGRI~$^{25}$ \\
53638.0 & $ 143.1800 \pm 0.1695 $ & INTEGRAL/ISGRI~$^{25}$ & 53641.0 & $ 143.0350 \pm 0.2270 $ & INTEGRAL/ISGRI~$^{25}$ \\
53644.0 & $ 143.0580 \pm 0.2625 $ & INTEGRAL/ISGRI~$^{25}$ & 53647.0 & $ 142.6760 \pm 0.7286 $ & INTEGRAL/ISGRI~$^{25}$ \\
53650.0 & $ 142.9760 \pm 0.3018 $ & INTEGRAL/ISGRI~$^{25}$ & 53653.0 & $ 143.0450 \pm 0.1820 $ & INTEGRAL/ISGRI~$^{25}$ \\
53656.0 & $ 143.1370 \pm 0.2403 $ & INTEGRAL/ISGRI~$^{25}$ & 53659.0 & $ 143.1670 \pm 0.2502 $ & INTEGRAL/ISGRI~$^{25}$ \\
53662.0 & $ 143.2370 \pm 0.1376 $ & INTEGRAL/ISGRI~$^{25}$ & 53665.0 & $ 143.2250 \pm 0.2702 $ & INTEGRAL/ISGRI~$^{25}$ \\
53668.0 & $ 143.0780 \pm 0.1641 $ & INTEGRAL/ISGRI~$^{25}$ & 53776.0 & $ 143.7540 \pm 0.4973 $ & INTEGRAL/ISGRI~$^{25}$ \\
53779.0 & $ 143.8440 \pm 0.1433 $ & INTEGRAL/ISGRI~$^{25}$ & 53782.0 & $ 143.8710 \pm 0.1307 $ & INTEGRAL/ISGRI~$^{25}$ \\
53785.0 & $ 143.8310 \pm 0.1705 $ & INTEGRAL/ISGRI~$^{25}$ & 53788.0 & $ 143.9010 \pm 0.0930 $ & INTEGRAL/ISGRI~$^{25}$ \\
53791.0 & $ 143.9800 \pm 0.0914 $ & INTEGRAL/ISGRI~$^{25}$ & 53794.0 & $ 144.0310 \pm 0.1641 $ & INTEGRAL/ISGRI~$^{25}$ \\
53797.0 & $ 144.0950 \pm 0.2588 $ & INTEGRAL/ISGRI~$^{25}$ & 53803.0 & $ 144.0920 \pm 0.1138 $ & INTEGRAL/ISGRI~$^{25}$ \\
53806.0 & $ 144.2470 \pm 0.1827 $ & INTEGRAL/ISGRI~$^{25}$ & 53809.0 & $ 144.0340 \pm 0.1718 $ & INTEGRAL/ISGRI~$^{25}$ \\
53812.0 & $ 144.1330 \pm 0.1196 $ & INTEGRAL/ISGRI~$^{25}$ & 53815.0 & $ 144.0920 \pm 0.1236 $ & INTEGRAL/ISGRI~$^{25}$ \\
53818.0 & $ 144.5760 \pm 0.2612 $ & INTEGRAL/ISGRI~$^{25}$ & 53821.0 & $ 144.2650 \pm 0.1599 $ & INTEGRAL/ISGRI~$^{25}$ \\
53830.0 & $ 144.2780 \pm 0.1662 $ & INTEGRAL/ISGRI~$^{25}$ & 53833.0 & $ 144.1490 \pm 0.1517 $ & INTEGRAL/ISGRI~$^{25}$ \\
53836.0 & $ 144.2310 \pm 0.2570 $ & INTEGRAL/ISGRI~$^{25}$ & 53839.0 & $ 144.2210 \pm 0.1294 $ & INTEGRAL/ISGRI~$^{25}$ \\
53842.0 & $ 144.4540 \pm 0.1796 $ & INTEGRAL/ISGRI~$^{25}$ & 53845.0 & $ 144.4040 \pm 0.0892 $ & INTEGRAL/ISGRI~$^{25}$ \\
53965.0 & $ 145.3250 \pm 0.2286 $ & INTEGRAL/ISGRI~$^{25}$ & 53968.0 & $ 145.5370 \pm 0.1574 $ & INTEGRAL/ISGRI~$^{25}$ \\
53974.0 & $ 145.4100 \pm 0.1504 $ & INTEGRAL/ISGRI~$^{25}$ & 53977.0 & $ 145.5250 \pm 0.1218 $ & INTEGRAL/ISGRI~$^{25}$ \\
53980.0 & $ 145.4940 \pm 0.1308 $ & INTEGRAL/ISGRI~$^{25}$ & 53983.0 & $ 145.5390 \pm 0.0991 $ & INTEGRAL/ISGRI~$^{25}$ \\
53986.0 & $ 145.5590 \pm 0.0952 $ & INTEGRAL/ISGRI~$^{25}$ & 53989.0 & $ 145.7140 \pm 0.0863 $ & INTEGRAL/ISGRI~$^{25}$ \\
53992.0 & $ 145.7020 \pm 0.0868 $ & INTEGRAL/ISGRI~$^{25}$ & 53998.0 & $ 145.7880 \pm 0.1160 $ & INTEGRAL/ISGRI~$^{25}$ \\
54001.0 & $ 145.6570 \pm 0.1365 $ & INTEGRAL/ISGRI~$^{25}$ & 54004.0 & $ 145.8051 \pm 0.1143 $ & INTEGRAL/ISGRI~$^{25}$ \\
54010.0 & $ 145.9100 \pm 0.0883 $ & INTEGRAL/ISGRI~$^{25}$ & 54013.0 & $ 145.8280 \pm 0.1115 $ & INTEGRAL/ISGRI~$^{25}$ \\
54022.0 & $ 145.8840 \pm 0.0907 $ & INTEGRAL/ISGRI~$^{25}$ & 54025.0 & $ 146.0170 \pm 0.0638 $ & INTEGRAL/ISGRI~$^{25}$ \\
54028.0 & $ 146.1190 \pm 0.0735 $ & INTEGRAL/ISGRI~$^{25}$ & 54031.0 & $ 146.0260 \pm 0.0850 $ & INTEGRAL/ISGRI~$^{25}$ \\
54034.0 & $ 146.0960 \pm 0.0748 $ & INTEGRAL/ISGRI~$^{25}$ & 54148.0 & $ 147.2450 \pm 0.0681 $ & INTEGRAL/ISGRI~$^{25}$ \\
54151.0 & $ 147.3313 \pm 0.0736 $ & INTEGRAL/ISGRI~$^{25}$ & 54157.0 & $ 147.2670 \pm 0.1061 $ & INTEGRAL/ISGRI~$^{25}$ \\
54160.0 & $ 147.2800 \pm 0.0755 $ & INTEGRAL/ISGRI~$^{25}$ & 54163.0 & $ 147.3790 \pm 0.0746 $ & INTEGRAL/ISGRI~$^{25}$ \\
54166.0 & $ 147.4210 \pm 0.0999 $ & INTEGRAL/ISGRI~$^{25}$ & 54169.0 & $ 147.4370 \pm 0.0758 $ & INTEGRAL/ISGRI~$^{25}$ \\
54175.0 & $ 147.5010 \pm 0.0742 $ & INTEGRAL/ISGRI~$^{25}$ & 54178.0 & $ 147.5084 \pm 0.0693 $ & INTEGRAL/ISGRI~$^{25}$ \\
54187.0 & $ 147.6360 \pm 0.0680 $ & INTEGRAL/ISGRI~$^{25}$ & 54193.0 & $ 147.7020 \pm 0.1296 $ & INTEGRAL/ISGRI~$^{25}$ \\
54196.0 & $ 147.7520 \pm 0.0915 $ & INTEGRAL/ISGRI~$^{25}$ & 54205.0 & $ 147.6560 \pm 0.5624 $ & INTEGRAL/ISGRI~$^{25}$ \\
54208.0 & $ 148.0240 \pm 0.1515 $ & INTEGRAL/ISGRI~$^{25}$ & 54211.0 & $ 147.8410 \pm 0.1414 $ & INTEGRAL/ISGRI~$^{25}$ \\
54214.0 & $ 147.9820 \pm 0.0789 $ & INTEGRAL/ISGRI~$^{25}$ & 54334.0 & $ 149.1250 \pm 0.1883 $ & INTEGRAL/ISGRI~$^{25}$ \\
54337.0 & $ 149.2490 \pm 0.1609 $ & INTEGRAL/ISGRI~$^{25}$ & 54340.0 & $ 149.2190 \pm 0.1271 $ & INTEGRAL/ISGRI~$^{25}$ \\
54346.0 & $ 149.2859 \pm 0.1262 $ & INTEGRAL/ISGRI~$^{25}$ & 54355.0 & $ 149.4520 \pm 0.1408 $ & INTEGRAL/ISGRI~$^{25}$ \\
54358.0 & $ 149.2110 \pm 0.1932 $ & INTEGRAL/ISGRI~$^{25}$ & 54361.0 & $ 149.7855 \pm 0.4506 $ & INTEGRAL/ISGRI~$^{25}$ \\
54364.0 & $ 149.5520 \pm 0.4772 $ & INTEGRAL/ISGRI~$^{25}$ & 54367.0 & $ 149.7460 \pm 0.2236 $ & INTEGRAL/ISGRI~$^{25}$ \\
54370.0 & $ 149.3710 \pm 0.1449 $ & INTEGRAL/ISGRI~$^{25}$ & 54379.0 & $ 149.6473 \pm 0.1996 $ & INTEGRAL/ISGRI~$^{25}$ \\
54388.0 & $ 149.7940 \pm 0.1169 $ & INTEGRAL/ISGRI~$^{25}$ & 54391.0 & $ 149.6390 \pm 0.1620 $ & INTEGRAL/ISGRI~$^{25}$ \\
54541.0 & $ 150.5900 \pm 0.3070 $ & INTEGRAL/ISGRI~$^{25}$ & 54704.2 & $ 152.7987 \pm 0.0017 $ & Fermi/GBM~$^{25}$ \\
54719.5 & $ 152.9342 \pm 0.0027 $ & Fermi/GBM~$^{25}$ & 54728.1 & $ 153.0133 \pm 0.0037 $ & Fermi/GBM~$^{25}$ \\
54736.1 & $ 153.0986 \pm 0.0024 $ & Fermi/GBM~$^{25}$ & 54744.1 & $ 153.1724 \pm 0.0018 $ & Fermi/GBM~$^{25}$ \\
54752.1 & $ 153.2497 \pm 0.0031 $ & Fermi/GBM~$^{25}$ & 54769.0 & $ 153.3928 \pm 0.0030 $ & Fermi/GBM~$^{25}$ \\
54775.9 & $ 153.4629 \pm 0.0016 $ & Fermi/GBM~$^{25}$ & 54784.0 & $ 153.5438 \pm 0.0012 $ & Fermi/GBM~$^{25}$ \\
54791.9 & $ 153.6298 \pm 0.0012 $ & Fermi/GBM~$^{25}$ & 54800.4 & $ 153.7102 \pm 0.0024 $ & Fermi/GBM~$^{25}$ \\
54808.1 & $ 153.7824 \pm 0.0025 $ & Fermi/GBM~$^{25}$ & 54815.7 & $ 153.8466 \pm 0.0021 $ & Fermi/GBM~$^{25}$ \\
54824.5 & $ 153.9263 \pm 0.0010 $ & Fermi/GBM~$^{25}$ & 54831.9 & $ 154.0113 \pm 0.0007 $ & Fermi/GBM~$^{25}$ \\
54840.0 & $ 154.1071 \pm 0.0007 $ & Fermi/GBM~$^{25}$ & 54847.9 & $ 154.1913 \pm 0.0011 $ & Fermi/GBM~$^{25}$ \\
54856.2 & $ 154.2690 \pm 0.0022 $ & Fermi/GBM~$^{25}$ & 54880.0 & $ 154.4594 \pm 0.0030 $ & Fermi/GBM~$^{25}$ \\
54887.9 & $ 154.5198 \pm 0.0023 $ & Fermi/GBM~$^{25}$ & 54904.7 & $ 154.6587 \pm 0.0032 $ & Fermi/GBM~$^{25}$ \\
54912.2 & $ 154.7181 \pm 0.0020 $ & Fermi/GBM~$^{25}$ & 54919.8 & $ 154.7824 \pm 0.0020 $ & Fermi/GBM~$^{25}$ \\
54927.9 & $ 154.8590 \pm 0.0024 $ & Fermi/GBM~$^{25}$ & 54936.3 & $ 154.9303 \pm 0.0016 $ & Fermi/GBM~$^{25}$ \\
54944.0 & $ 154.9999 \pm 0.0016 $ & Fermi/GBM~$^{25}$ & 54952.1 & $ 155.0761 \pm 0.0021 $ & Fermi/GBM~$^{25}$ \\
54959.5 & $ 155.1347 \pm 0.0023 $ & Fermi/GBM~$^{25}$ & 54968.0 & $ 155.2070 \pm 0.0019 $ & Fermi/GBM~$^{25}$ \\
54975.9 & $ 155.2761 \pm 0.0021 $ & Fermi/GBM~$^{25}$ & 54984.0 & $ 155.3446 \pm 0.0014 $ & Fermi/GBM~$^{25}$ \\
54992.0 & $ 155.4194 \pm 0.0010 $ & Fermi/GBM~$^{25}$ & 55000.0 & $ 155.4976 \pm 0.0010 $ & Fermi/GBM~$^{25}$ \\
55007.9 & $ 155.5781 \pm 0.0010 $ & Fermi/GBM~$^{25}$ & 55016.1 & $ 155.6549 \pm 0.0019 $ & Fermi/GBM~$^{25}$ \\
55023.9 & $ 155.7235 \pm 0.0018 $ & Fermi/GBM~$^{25}$ & 55031.9 & $ 155.7919 \pm 0.0023 $ & Fermi/GBM~$^{25}$ \\
55040.0 & $ 155.8628 \pm 0.0026 $ & Fermi/GBM~$^{25}$ & 55047.9 & $ 155.9249 \pm 0.0016 $ & Fermi/GBM~$^{25}$ \\
55056.0 & $ 156.0042 \pm 0.0009 $ & Fermi/GBM~$^{25}$ & 55063.8 & $ 156.0889 \pm 0.0007 $ & Fermi/GBM~$^{25}$ \\
55072.3 & $ 156.1731 \pm 0.0015 $ & Fermi/GBM~$^{25}$ & 55079.8 & $ 156.2443 \pm 0.0016 $ & Fermi/GBM~$^{25}$ \\
55087.9 & $ 156.3116 \pm 0.0021 $ & Fermi/GBM~$^{25}$ & 55120.0 & $ 156.5714 \pm 0.0011 $ & Fermi/GBM~$^{25}$ \\
55128.1 & $ 156.6515 \pm 0.0007 $ & Fermi/GBM~$^{25}$ & 55135.9 & $ 156.7381 \pm 0.0008 $ & Fermi/GBM~$^{25}$ \\
55143.9 & $ 156.8283 \pm 0.0007 $ & Fermi/GBM~$^{25}$ & 55151.9 & $ 156.9130 \pm 0.0009 $ & Fermi/GBM~$^{25}$ \\
55159.9 & $ 157.0009 \pm 0.0006 $ & Fermi/GBM~$^{25}$ & 55168.0 & $ 157.0931 \pm 0.0005 $ & Fermi/GBM~$^{25}$ \\
55176.0 & $ 157.1903 \pm 0.0005 $ & Fermi/GBM~$^{25}$ & 55184.2 & $ 157.2875 \pm 0.0006 $ & Fermi/GBM~$^{25}$ \\
55191.9 & $ 157.3714 \pm 0.0009 $ & Fermi/GBM~$^{25}$ & 55200.0 & $ 157.4465 \pm 0.0014 $ & Fermi/GBM~$^{25}$ \\
55207.9 & $ 157.5144 \pm 0.0022 $ & Fermi/GBM~$^{25}$ & 55224.0 & $ 157.6355 \pm 0.0023 $ & Fermi/GBM~$^{25}$ \\
55232.0 & $ 157.7167 \pm 0.0006 $ & Fermi/GBM~$^{25}$ & 55237.5 & $ 157.7838 \pm 0.0019 $ & Fermi/GBM~$^{25}$ \\
\hline
\hline

\end{longtable}

\tablebib{
(1)~\citet{GX71Lewin}; (2)~\citet{GX76White}; (3)~\citet{GX80Koo}; (4)~\citet{GX76Becker}; (5)~\citet{GX81Doty}; (6)~Becker \& White in \citet{GX86Cut}; (7)~\citet{GX80Strick}; (8)~\citet{GX81Coe}; (9)~\citet{GX82Kend}; (10)~\citet{GX85Elsner}; (11)~\citet{GX82Ric}; (12)~\citet{GX87Jayan} (Three probable periods centred at 108.8, 112.8, and 121.2 s were selected in this paper and they decided in favour of 120.6 s, but the trend indicates the best period is 108.8 s.); (13)~\citet{GX88Dam}; (14)~\citet{GX89Leahy}; (15)~\citet{GX93Green}; (16)~\citet{GX88Mak}; (17)~\citet{GX91Mony}; (18)~\citet{GX89Dotani}; (19)~\citet{GX90Sakao}; (20)~\citet{GX94Luto}; (21)~\citet{GX93Lau}; (22)~\citet{GX97Chack}; (23)~\citet{GX98David}; (24)~\citet{GX05Naik}; (25)~This work; (26)~\citet{GX04Cui}; (27)~\citet{GX05Paul}; (28)~\citet{GX07Ferrigno}, (29)~BATSE Public Data Archive.}
}

\end{document}